\documentclass[aps,prx,reprint,amsmath,amssymb,superscriptaddress,nofootinbib,noeprint,longbibliography,floatfix,nobibnotes]{revtex4-2}
\pdfoutput=1

\usepackage{dsfont}
\usepackage{newtxtext}
\usepackage{bbold}
\usepackage{braket}
\usepackage[dvipsnames]{xcolor}
\usepackage{microtype}
\usepackage{placeins}
\usepackage{amsthm}
\usepackage{nameref}
\usepackage{mathtools}
\usepackage{CJKutf8}

\newtheorem{theorem}{Theorem}
\newtheorem{corollary}{Corollary}[theorem]
\newtheorem{lemma}{Lemma}
\newtheorem{definition}{Definition}

\definecolor{myciteColor}{rgb}{0.0,0.5,0.23}
\usepackage[colorlinks=true,citecolor=myciteColor,linkcolor=myciteColor,urlcolor=myciteColor]{hyperref}

\usepackage{orcidlink}

\usepackage{tcolorbox}

\begin{abstract}
\normalfont Analog quantum simulators have provided key insights into quantum many-body dynamics. However, in such systems, both coherent and incoherent errors limit their scalability, hindering simulations in regimes that challenge classical simulations. In this work, we introduce an error mitigation technique that addresses and effectively suppresses a key source of error in leading simulator platforms: shot-to-shot fluctuations in the parameters for the Hamiltonian governing the system dynamics. We rigorously prove that amplifying this shot-to-shot noise and extrapolating to the zero-noise limit recovers noiseless results for realistic noise distributions. Experimentally, we demonstrate this technique on a 27-ion trapped-ion quantum simulator, extending the two-qubit exchange oscillation lifetime threefold. Numerically, we predict a significant enhancement in the effective many-body coherence time for Rydberg atom arrays under realistic conditions. Our scheme provides a possible route towards extending the effective coherence time in analog quantum experiments, enabling deeper explorations of quantum many-body dynamics.
\end{abstract}

\begin{document}
\begin{CJK}{UTF8}{gbsn}
\title{Error mitigation of shot-to-shot fluctuations in analog quantum simulators}

\author{Thomas Steckmann\,\orcidlink{0000-0001-6012-2948}}
\affiliation{Joint Center for Quantum Information and Computer Science,
University of Maryland and NIST, College Park, Maryland 20742, USA}
\affiliation{Joint Quantum Institute,
University of Maryland and NIST, College Park, Maryland 20742, USA}

\author{De Luo\,\orcidlink{0000-0002-4842-0990}}
\affiliation{Duke Quantum Center, Department of Physics, and Department of Electrical and Computer Engineering, Duke University, Durham, NC 27701 USA}

\author{Yu-Xin Wang (王语馨)\,\orcidlink{0000-0003-2848-1216}}
\affiliation{Joint Center for Quantum Information and Computer Science,
University of Maryland and NIST, College Park, Maryland 20742, USA}

\author{Sean R. Muleady\,\orcidlink{0000-0002-5005-3763
}}
\affiliation{Joint Center for Quantum Information and Computer Science,
University of Maryland and NIST, College Park, Maryland 20742, USA}
\affiliation{Joint Quantum Institute,
University of Maryland and NIST, College Park, Maryland 20742, USA}

\author{Alireza Seif\,\orcidlink{0000-0001-5419-5999}}
\affiliation{IBM Quantum, IBM T.J. Watson Research Center, Yorktown Heights, NY 10598, USA}

\author{Christopher Monroe\,\orcidlink{0000-0003-0551-3713}}
\affiliation{Duke Quantum Center, Department of Physics, and Department of Electrical and Computer Engineering, Duke University, Durham, NC 27701 USA}

\author{Michael J. Gullans\,\orcidlink{0000-0003-3974-2987}}
\affiliation{Joint Center for Quantum Information and Computer Science,
University of Maryland and NIST, College Park, Maryland 20742, USA}

\author{Alexey V. Gorshkov\, \orcidlink{0000-0003-0509-3421}}
\affiliation{Joint Center for Quantum Information and Computer Science,
University of Maryland and NIST, College Park, Maryland 20742, USA}
\affiliation{Joint Quantum Institute,
University of Maryland and NIST, College Park, Maryland 20742, USA}

\author{Or Katz\,\orcidlink{0000-0001-7634-1993}}
\affiliation{School of Applied and Engineering Physics, Cornell University, Ithaca, NY 14853.}

\author{Alexander Schuckert\,\orcidlink{0000-0002-9969-7391}}
\affiliation{Joint Center for Quantum Information and Computer Science,
University of Maryland and NIST, College Park, Maryland 20742, USA}
\affiliation{Joint Quantum Institute,
University of Maryland and NIST, College Park, Maryland 20742, USA}
\date{\today}
\maketitle
\end{CJK}

\section{Introduction}
Analog quantum simulators offer unique advantages compared to digital quantum computers for studying certain problems in many-body quantum dynamics~\cite{Bluvstein2021, Daley2022, choi_robust_2020}. Fewer control requirements enable scaling to comparably large system sizes, approaching scales which are challenging
to describe using exact classical methods~\cite{Aspuru-Guzik2012, zhang2017, Scholl2021, semeghini2021}. Additionally, for Hamiltonians within the class that may be emulated on a particular system---for example, interacting bosons or fermions, in both continuous space and on a lattice~\cite{greiner2002,jordens2008,trotzky2012,cheuk_observation_2016,mazurenko_cold-atom_2017,nichols2019,carroll_observation_2024}, or spin models featuring Ising~\cite{labuhn_tunable_2016,zhang2017,bernien_probing_2017,semeghini2021,guo2024,schuckert2025} or XY~\cite{yan_observation_2013,Feng2023,katz2024observing,joshi_observing_2022} interactions---long-time evolution can be performed without the need for costly Trotterization methods.

However, analog quantum simulators are subject to limitations in the accuracy of the engineered interactions, and despite some specialized quantum error correction protocols~\cite{Bacon2006,Jordan2006,pudenz2014,marvian2014,marvian2017,cao2024}, the absence of a generally applicable, scalable concept for fault tolerance renders scaling challenging. As an alternative, their reliability can be improved through the study and mitigation of leading error sources, which are often fluctuations of the programmable control fields used to design the system Hamiltonian. When these fluctuations generate unwanted terms not part of the target Hamiltonian, they can be removed by dynamical decoupling~\cite{Suter2016,Morong2023,choi_robust_2020}; however, fluctuations in the parameters of the desired Hamiltonian are, to our knowledge, not amenable to existing dynamical decoupling schemes. Since these fluctuations are often ``colored'' noise (e.g.~pink noise scaling as $1/f$), their magnitude increases with the observational time scale.
The strongest fluctuations of the Hamiltonian parameters therefore occur on a time scale such that they are constant over the course of many repetitions of an experiment: they are therefore effectively static and may be learned and accounted for via sufficiently frequent calibration. The next-leading time scale of relevance is drift occurring over the course of a single experiment, but which remains static on the time scale of a single repetition (shot) of an experiment. These \emph{shot-to-shot fluctuations}, also known as quasi-static errors, cannot be learned except on average and thus cannot be accounted for by calibration. Indeed, such fluctuations are often the leading source of error in analog quantum simulators, outweighing Markovian errors, in particular for ion-traps with individual ion addressing~\cite{cetina2022} and Rydberg atom experiments~\cite{shaw2024}. In addition, these shot-to-shot fluctuations sometimes become stronger as the system size increases, directly limiting the geometry and practical scale of experiments. For instance, in trapped-ion simulators with individual addressing, $1/f$ noise induces an error in two-body entangling interactions scaling as $N^6$ with the number of ions $N$~\cite{cetina2022}. While this noise can be suppressed with sympathetic cooling, its implementation often requires multi-species trapping~\cite{cetina2022,pino2021demonstration} or exquisite control of the internal state~\cite{allcock2021omg}.

Independently---in the context of noisy quantum circuits---the idea of mitigating errors using zero-noise extrapolation (ZNE) has been introduced~\cite{li2017,temme2017} and successfully applied in many experiments~\cite{Kandala2019,Kim2023,hemery2024}: by carefully amplifying the noise, one hopes to extrapolate observables to the zero noise limit. However, ZNE relies on scaling specific error parameters independent of the target dynamics. How to identify and scale such a parameter in the setting of shot-to-shot fluctuations and analog controls has not been considered and is not compatible with existing ZNE implementations. Shot-to-shot errors therefore constitute a fundamental barrier to scaling analog simulation to longer times and larger system sizes, with no technique available to mitigate these errors without imposing extra hardware constraints. 

Here, we overcome this barrier to scaling analog quantum simulation experiments by introducing ZNE for random shot-to-shot fluctuations of a single Hamiltonian parameter. To show that the idea of ZNE is applicable to shot-to-shot noise, we first prove that observables are smooth functions of the noise for sufficiently weak noise and then prove a rigorous error bound on the remaining error after ZNE. We then illustrate the efficiency of this method by implementing it experimentally in a trapped-ion system, showing a large improvement of the effective coherence time, demonstrating a route towards mitigating this leading error source. In addition, we show numerically that our methods yield similarly large improvements in Rydberg-atom simulators. While inspired by digital ZNE, our analog ZNE scheme is distinct: it leverages the hardware-level control available in analog simulators to tailor the mitigation technique to specific sources of noise, enabling an implementation in analog simulators without any additional control requirements. 

The remainder of this paper is organized as follows. In section~\ref{sct:def}, we formalize the error model considered in this work. Then, in section~\ref{sct:proof}, we prove a rigorous error bound on the applicability of our scheme to mitigating such noise. We then apply our result to two leading analog quantum simulation platforms: Rydberg atom arrays and trapped ions. For the former, in section~\ref{sct:ryd}, we numerically show that the lifetime of emergent many-body oscillations---indicating the presence of quantum many-body scars---can be extended by our analog ZNE scheme. For the latter, in section~\ref{sct:ions}, we experimentally implement our protocol in a  trapped ion chain with individual addressing, showing its feasibility by extending the lifetime of two-qubit exchange oscillations. In section~\ref{sec:discussion}, we present discussion and outlook. 

\section{Theoretical framework for analog error mitigation}
\subsection{Definition of shot-to-shot noise}~\label{sct:def}
We begin by formalizing and illustrating our definition of shot-to-shot, or quasi-static, noise while providing a sufficient condition for mitigable errors under this model. Motivated by the fact that control beams with a small number of fluctuating parameters are the primary source of shot-to-shot noise in individually addressed trapped-ion~\cite{cetina2022} and Rydberg-atom experiments~\cite{shaw2024}, we focus on perturbations arising from fluctuations of a single Hamiltonian control parameter.

\begin{tcolorbox}[colback=LimeGreen!20, colframe=gray!100,rounded corners=all,boxrule=0.5mm, width=\columnwidth]
\begin{definition}[Shot-to-shot noise]
Shot-to-shot noise is a coherent perturbation $\hat{V}$ of the ideal Hamiltonian $\hat{H}_0$ with a dimensionless noise parameter $\delta$ that varies in each repetition (i.e.~shot) of the experiment. Specifically, the Hamiltonian in each shot is given by
\begin{equation}
\hat{H} = \hat{H}_{0} +\delta \hat{V},\label{eq:defshot}
\end{equation}
with $\delta$ drawn independently from a probability distribution $\mathcal{D}_\theta$ parametrized by $\theta \ge 0$. The distribution is defined such that $\theta=0$ corresponds to $\delta=0$ in every experimental shot.\label{def:shot-to-shot}
\end{definition}
\end{tcolorbox}
This definition includes, but is not restricted to, the case of multiplicative fluctuations in a Hamiltonian component,
$\hat{H} = \hat H_1 + (1+\delta)\hat H_2,$
which is translated into the form in Eq.~\eqref{eq:defshot} with $\hat V=\hat H_2$ and $\hat H_0=\hat H_1 + \hat H_2$. We do not impose constraints on $\hat{V}$, meaning it need not commute with $\hat{H}_0$ and may represent multi-body terms. For example, $\hat{V}$ can represent an Ising term $\hat{V} \propto \sum_{ij} J_{ij} \hat \sigma^x_i \hat \sigma^x_j$, with $\hat \sigma_i^\alpha$ ($\alpha = x,y,z$) denoting a Pauli matrix acting on the $i^\mathrm{th}$ qubit. We note that while multiplicative fluctuations may allow relative scaling, which can be exploited to perform ZNE when the dynamics follow a known scaling law~\cite{raymond_amin2025, guo2025a}, in general such errors are not amenable to pulse-stretching error mitigation~\cite{maupin2024}.

A simple example of such noise is the fluctuation of the single-qubit Rabi frequency around a target value $\Omega$, e.g.~due to amplitude fluctuations of the drive field. In the frame rotating at the drive frequency, the dynamics are described by $\hat H_0+\delta\hat V = (1+\delta) \Omega \hat \sigma^x$, where $\delta \Omega$ quantifies the noise strength, and we set the reduced Planck constant $\hbar=1$ without loss of generality. We define $\ket{\uparrow}$ ($\ket{\downarrow}$) as the $+1$ ($-1$) eigenstate of $\hat \sigma ^z$. Starting in $\ket{\downarrow}$, the ideal dynamics with $\delta=0$ lead to undamped oscillations of the transfer probability to $\ket{\uparrow}$:
\begin{equation}
    P_{\uparrow,\Omega}(t)=\frac{1}{2}\left(1-\cos(2\Omega t)\right).
\end{equation}
When $\delta$ fluctuates between different shots of the experiment, the oscillations of the shot-averaged transfer probability become damped, as illustrated in 
\begin{figure*}[t]
    \centering
    \includegraphics[width=2\columnwidth]{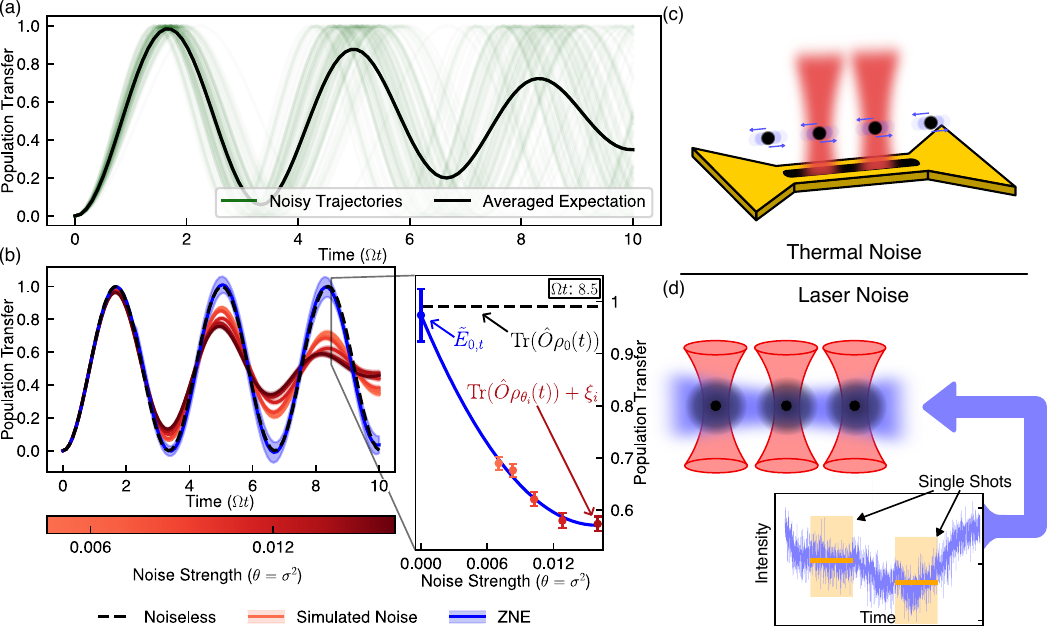}
    \caption{\textbf{Shot-to-shot noise: illustration and physical origin in experiments.} (a) Population transfer of single-qubit Rabi oscillations as a function of time starting from initial state $\ket{\downarrow}$ for varying Rabi frequency $\Omega(1+\delta)$ (green lines), where $\delta$ is sampled from a Gaussian distribution with zero mean and minimum variance $\theta_0 = \sigma_0^2 = 0.0064$. The average is shown in black. (b) Numerical demonstration of analog ZNE applied to the example in (a). Red lines with increasing darkness represent averages of $600$ noisy trajectories with increasing noise strength (variance) of the fluctuations. The dashed black line is the exact result without fluctuations. The blue line is the result from the ZNE, with faint bands indicating the error from the polynomial fit. (inset) The red dots show the population transfer at time $t$ such that $\Omega t = 8.5$ plotted against the noise strength, with error bars given by the standard deviation of averaging over $600$ trajectories. The blue curve is a second-order fit as determined by a leave-one-out cross validation (LOOCV) method, with the error bar of the ZNE estimate $\tilde{E}_0$ (blue point) determined by the error of the fit. $\hat{\rho}_{\theta_i}(t)$ is the density matrix in the presence of shot noise with strength $\theta_i$, $\xi_i$ is the projection noise, and $\hat{\rho}_0(t)$ the density matrix without shot noise. (c) Axial center-of-mass motion in a linear ion chain leads to shot-to-shot fluctuations of the Rabi frequency due to ion motion with respect to individual addressing Raman beams. (d) Slow laser-intensity fluctuations in the Rydberg laser arising from low-frequency noise lead to shot-to-shot fluctuations of the ground-Rydberg Rabi frequency in an array of neutral atoms.}
    \label{fig:1}
\end{figure*}
Fig.~\ref{fig:1}(a).

Returning to the consideration of generic shot-to-shot noises, to prove our theorem on analog ZNE, we need to impose some constraints on the error distribution $\mathcal{D}_\theta$, which hold for typical physical noise sources. 

\begin{tcolorbox}[colback=LimeGreen!20, colframe=gray!100,rounded corners=all,boxrule=0.5mm, width=\columnwidth]
\begin{definition}[Mitigable distribution]
    We call a distribution $\mathcal{D}_\theta$ mitigable with respect to $\theta$ if, for all $k\in \mathbb{Z}^+$, the $k^\mathrm{th}$ moment $\mathbb{E}_\theta[\delta^k] = \int \delta^k \mathcal{D}_{\theta}(\delta) \mathrm{d}\delta$ can be expressed as a polynomial $p_k(\theta)$ with $\deg(p_k(\theta)) \leq k$. \label{lemma:mitigable-distributions}
\end{definition}
\end{tcolorbox}

An example for a mitigable distribution is the Gaussian distribution with zero mean, 
\begin{equation}
    \mathcal{D}_{\sigma^2}(\delta)=\frac{1}{\sqrt{2\pi}\sigma}\exp\left(-\frac{\delta^2}{2\sigma^2}\right),
    \label{eq:Gauss_def}
\end{equation}
as its $k^\mathrm{th}$ moments can be expressed as a polynomial in terms of its variance $\theta=\sigma^2$ with degree less than or equal to $k$. Assuming such Gaussian-distributed amplitude fluctuations in our previous Rabi-oscillation example, we analytically evaluate the averaged transfer probability $\mathbb{E}_{\sigma^2}\left[P_\uparrow(t)\right]=\bar{P}_{\uparrow}(t)$ in the presence of shot-to-shot fluctuations:
\begin{align}
   \bar P_\uparrow(t) &= \int_{-\infty}^\infty d\delta\,\mathcal{D}_{\sigma^2}(\delta)  P_{\uparrow,\Omega(1+\delta)}(t)\\
    &= \frac{1}{2}\left(1-e^{-2\sigma^2\Omega^2 t^2} \cos(2\Omega t) \right).
    \label{eq:Gauss}
\end{align}
The oscillations acquire a decaying envelope which, unlike the dynamics under Markovian dephasing noise, is Gaussian rather than exponential, and the initial decay is hence quadratic instead of linear in time.

The expression in Eq.~\eqref{eq:Gauss} inspires the idea of using a version of ZNE~\cite{temme2017}: if we were able to controllably enhance $\sigma^2$, we could record $\bar P_\uparrow(t)$ as a function of $\sigma^2$ and then extrapolate to $\sigma^2\rightarrow 0$. To show that this is indeed possible, we assume that the noise is sufficiently small such that $\sigma\Omega t\ll 1$, and expand the exponential:
\begin{align}
    \bar P_\uparrow(t)= P_\uparrow(t)-2(\sigma\Omega t)^2\cos(2\Omega t)+\mathcal{O}\left((\sigma\Omega t)^4\right).
\end{align}
Thus, for small $(\sigma\Omega t)^2$, we can recover $P_\uparrow(t)$ for fixed $t$ by recording $\bar P_\uparrow(t)$ as a function of $\sigma^2$, fitting to a linear function in $\sigma^2$, and extrapolating to the value of $\bar P_\uparrow(t)$ at $\sigma^2=0$. We implement this procedure numerically in Fig.~\ref{fig:1}(b), showing that undamped Rabi oscillations can be recovered even when $\sigma_0^2 (\Omega t)^2\approx 0.37 $, with $\sigma_0^2$ denoting the smallest noise strength. 
\subsection{Proof of zero-noise extrapolation}~\label{sct:proof}

How generally can ZNE be applied to shot-to-shot noise? In this section, we answer this question with our main theorem---proved in Appendix~\ref{app:proofs}---for the method of Richardson extrapolation and discuss the key differences between our method and existing techniques. 

Consider a mitigable distribution $\mathcal{D}_\theta$ and $r+1$ distinct values of $\theta$ labeled $\{\theta_{0},\dots \theta_{r}\}$. For each $\theta_i$, and given some observable $\hat{O}$  evaluated at evolution time $t$, $N$ measurements on the quantum simulator produce a noisy estimator \begin{equation}
    \tilde{E}_{\theta_{i},t}= \mathrm{Tr} (\hat{O}\hat\rho_{\theta_{i}}(t)) + \xi_{i}.
\end{equation}
$\tilde{E}_{\theta_{i},t}$ contains two sources of noise. On the one hand, it has the usual quantum projection noise $\xi_i$, which in the limit of large $N$ is normally distributed with variance $\nu_{i}^{2}$. On the other hand, there is shot-to-shot noise, which we absorb into the density matrix $\hat{\rho}_{\theta}(t)$ by averaging over the distribution $\mathcal{D}_{\theta}$,
\begin{align}
    \hat{\rho}_{\theta}(t) =
    \mathbb{E}_{\theta} \left[  e^{-it(\hat{H}_0 + \delta \hat{V})} \hat{\rho}(0)e^{it(\hat{H}_0 + \delta \hat{V})}\right],
\end{align}
where $\hat{\rho}(0)$ is the initial density matrix.

For ZNE, we then interpolate the results of these $r+1$ estimators. To get a well-defined bound on the error of the ZNE estimator, we use Richardson extrapolation in our Theorem (in practice, we use other techniques, see below). The estimator is given by the unique interpolating $r^\mathrm{th}$-order polynomial with respect to $\theta$ evaluated at $\theta=0$,  
\begin{equation}
\tilde{E}_{0,t}=\sum_{i=0}^r\gamma_{i}\tilde{E}_{\theta_{i},t}.
\end{equation}
Here, $\gamma_i= \prod_{k\neq i} \frac{\theta_k}{\theta_k-\theta_i}$ are the Lagrange basis polynomials evaluated at $\theta=0$~\cite{temme2017, Krebsbach2022}. Due to a multiplicative effect on projection error, the final Richardson extrapolation estimator is highly sensitive to these values. For example, with five linearly spaced nodes between a minimum value of $\gamma_0 =\theta_{0}$ and $\gamma_4 =2\theta_{0}$, $\sum_{i=0}^4 |\gamma_i| = 769\theta_0$, while for ten nodes between $\gamma_0 = \theta_{0}$ and $\gamma_9=2\theta_{0}$ this summation increases to $16,807,935\theta_0$. This instability aligns with Runge's phenomenon, and has been studied analytically in the context of ZNE in Refs.~\cite{Krebsbach2022, mohammadipour2025b}.

We are now in a position to bound the error of the ZNE estimator. This constitutes our main theorem:
\begin{tcolorbox}
    \begin{theorem}[Analog ZNE of Shot-to-Shot Noise]
Consider a set of $r+1$ estimators (each averaged over a large number of shots) on an analog quantum simulator subject to shot-to-shot noise drawn from a mitigable distribution $\mathcal{D}_{\theta}$, resulting in pairs of noise strengths and noisy estimators, $\{(\theta_{0}, \tilde{E}_{\theta_{0},t}), \dots, (\theta_{r}, \tilde{E}_{\theta_{r},t})\}$. We assume these are ordered by the value of $\theta_i$, and that $\tilde{E}_{\theta_{0},t}$ has projection noise distributed with variance $\nu_{i}^2$. Then, with probability $1-\epsilon$, the result of ZNE, via the Richardson extrapolator and absent additional information about $\mathcal{D}_\theta$, has an error bounded by
\begin{align}
    |\tilde{E}_{0,t}-\mathrm{Tr}&(\hat{O}\hat\rho_0(t))| \nonumber \\ &\leq \sum_{i=0}^{r}|\gamma_{i}|\left( |R_{r+1}( \mathcal{D}_{\theta_i},t,\hat{V})| + c_i \right),\label{eq:main-theorem-statement}
\end{align}
where the remainder error term $R$ is bounded by the following expression set by moments of the distribution $\mathcal{D}_{\theta},$
\begin{align}
    |R_{r+1}(\mathcal{D}_{\theta_i}, t, \hat{V})| &\leq \|\hat{O}\|_{\infty}\frac{(2t)^{r+1}\|\hat{V}\|_{\infty}^{r+1}\mathbb{E}_{\theta_i}\left[ |\delta^{r+1}|\right]}{(r+1)!},\label{eq:remainder-term-theorem}
\end{align}
and $c_i = \sqrt{ 2 }\nu_i \mathrm{erf}^{-1}(1-\epsilon)$ bounds the quantum projection noise in the large-shot-number limit. $\|\hat A \|_\infty$ is the spectral norm equal to the largest eigenvalue norm of $\hat A$.\label{thm:main}
    \end{theorem}
\end{tcolorbox}

Theorem \ref{thm:main} establishes the distribution parameter $\theta$ as a viable extrapolation axis for ZNE while providing an upper bound on the error of the resulting ZNE estimator dependent on simulation time ($t$), the moments of the noise distribution ($\mathbb{E}_{\theta_i}\left[ |\delta^{r+1}|\right]$), the perturbation operator ($\|\hat V\|_{\infty}$), projection noise ($\nu _{i}$), and the parameters of the Lagrange basis polynomials used to construct the Richardson estimator ($\gamma_i$). The error $R$ (Eq.~\eqref{eq:remainder-term-theorem}) arises from the  $r^\mathrm{th}$-order polynomial approximation in the absence of projection noise, and encodes the dependence of the error on the noise distribution $\mathcal{D}$: for large values $r$, the denominator scales as $(r+1)!$ and competes with the scaling of moments of $\mathcal{D}$. As we will demonstrate, for Gaussian distributed noise the remainder converges to zero for large $r$ and arbitrary noise strength, with moments scaling as $r!! \leq r!$ (Cor.~\ref{cor:ZNE-Normal}); for thermal distributions (Cor.~\ref{cor:thermal-ZNE}) these moments scale as $r!$ and $R$ only converges to zero for sufficiently small noise strengths. The ZNE estimator, $\tilde E_{0,t}$, has an error from the ideal estimator due to projection noise. The failure rate, $\epsilon$, in our bound arises from truncating the tails of the distribution from which the specific, and unknown, values of $\xi_i$ are sampled; in the case that an experimental estimator is outside the predefined error tolerance from projection noise, $c_i$, the inequality Eq.~\eqref{eq:main-theorem-statement} will not hold, but will hold for some other $c_i'$ corresponding to a larger error tolerance. 

While Richardson extrapolation offers a useful theoretical tool in our proof (it  minimizes the remainder error $R$ arising from the $r^\mathrm{th}$-order polynomial approximation in the absence of projection noise), practical implementations of ZNE on a quantum device operate in a regime of finite sampling, leading to a trade-off between projection noise and the remainder error. To avoid the instability of Richardson extrapolation arising from overfitting uncertain estimators with high-order polynomials~\cite{Krebsbach2022, Giurgica-Tiron2020}, we employ least-squares fitting and a leave-one-out cross-validation (LOOCV) method to select the polynomial order. This approach, detailed in Appendix~\ref{appendix: LOOCV}, balances the reduced truncation remainder $R$ for larger fit orders while maintaining the averaging effect of the least-square regression by selecting an appropriate polynomial order that avoids overfitting. However, while our LOOCV method performs well in practice, it is challenging to provide analytical bounds on the performance compared to Richardson extrapolation. 

Theorem~\ref{thm:main} has key implications for error mitigation in analog simulators. First, it extends the current scope of noise models for which ZNE may be applied. Existing methods for ZNE in an analog setting focus on dynamics modeled by a quantum master equation~\cite{temme2017}
\begin{align}
    \frac{\mathrm{d}\hat \rho}{\mathrm{d}t} = -i[\hat{H}, \hat\rho(t)] + \lambda  \mathcal{L}(\hat\rho(t)) \label{eq:temme-zne-model},
\end{align}
where $\mathcal{L}$ is a super-operator representing the error term, and can take on a variety of forms, while $\lambda$ serves as the parameter that is extrapolated to zero and must be amplified in a controlled way independent of the terms $\mathcal{L}$ and $\hat{H}$. However, Eq.~\eqref{eq:temme-zne-model} does not capture our model of shot-to-shot noise as the latter cannot be generated by a Markovian dissipative process. Thus, Theorem~\ref{thm:main} extends ZNE to a new setting, where $\theta$ serves as a valid extrapolation parameter and the target of experimental amplification. As we will introduce in sections~\ref{sct:ryd} and~\ref{sct:ions}, the physical error models associated with shot-to-shot noise provide experimental access to the distribution parameter such that it can be amplified and measured in a manner that neither requires additional control hardware nor directly modifies the target evolution; this is in contrast to existing methods relying on pulse-stretching~\cite{raymond_amin2025, maupin2024, guo2025a}. For maximum applicability in existing setups, our method does not rely on particular capabilities such as access to addressable single qubit gates during the experiment~\cite{sun2021, Kim2023}, access to time-reversed evolutions~\cite{meher2024, shaffer2021}, or discretized evolution under gates~\cite{Giurgica-Tiron2020, dumitrescu2018, He2020}.

\section{Applications}

In the remainder of our work, we show how Theorem~\ref{thm:main} can be applied to Rydberg atom and trapped-ion experiments that are subject to shot-to-shot noise with different types of error distributions $\mathcal{D}_\theta$. We first start with a numerical study of ZNE in Rydberg atoms coupled to a simpler Gaussian-distributed noise. We then move on to experimentally demonstrate ZNE of shot-to-shot noise in trapped ions.

\subsection{Gaussian laser fluctuations in Rydberg-interacting neutral atom arrays: extending the lifetime of many-body scars}\label{sct:ryd}

The signatures of the dynamical phenomena we aim to study using analog quantum simulators may be masked by the effects of shot-to-shot noise in the systems. Here, we apply Theorem~\ref{thm:main} to the specific case of Gaussian-distributed noise introduced in Eq.~\eqref{eq:Gauss_def} and demonstrate ZNE as a means of prolonging the lifetime of quantum many-body scars in neutral-atom arrays. While Corollary~\ref{cor:ZNE-Normal} applies to generic perturbations $\hat V$, in the context of Rydberg-interactions we primarily consider the case of $\hat V \propto \Omega$, such as in Eq.~\eqref{eq:Gauss} with $\|\hat V\|_\infty=\Omega$. By explicitly computing the moments of the Gaussian distribution, we obtain the following corollary, proven in Appendix~\ref{app:proofs}: 
\begin{tcolorbox}[colback=Maroon!10, colframe=gray!100,rounded corners=all,boxrule=0.5mm, width=\columnwidth]
\begin{corollary}[Analog ZNE for Gaussian-distributed noise]
    The zero-centered normal distribution defined in Eq.~\eqref{eq:Gauss_def} is mitigable and permits ZNE along the $\sigma^2$ axis with the remainder of Richardson extrapolation on $r+1$ experimental estimators given by \begin{equation}
        |R_{2(r+1)} (\mathcal{D}_{\sigma^2}, t, \hat V)| \leq \| \hat{O}\|_{\infty} \frac
        {\left(\sigma^{2}(2t)^2\|\hat{V}\|_\infty^2\right)^{r+1}}
        {(2r+2)!!}.
        \label{eq:gaussian-remainder}
    \end{equation}
    \label{cor:ZNE-Normal}
\end{corollary}
\end{tcolorbox}
This corollary shows that the Taylor series of $\mathrm{Tr}(\hat O \hat \rho_{\sigma^2} (t))$ with respect to $\sigma^2$ has an infinite radius of convergence, i.e.~the truncation error can be made arbitrarily small for sufficiently large $r$. This is because the factorial in the denominator grows faster in $r$ than the exponential in the numerator. Hence, in principle, ZNE works for any $t$ and any smallest noise strength $\sigma^2_0$. However, due to the practical considerations of Richardson instability and sampling overhead in suppressing the sampling error $\nu$, ZNE is typically performed by fitting to low-order polynomials or families of exponential functions~\cite{Giurgica-Tiron2020,Kim2023, Majumdar2023, Cai2021}. This in turn limits the usefulness of the guaranteed convergence at large values of $r$.

Additionally, due to the odd moments of the normal distribution evaluating to zero, the variance of the distribution $\sigma^2$ is used as the extrapolation parameter rather than the standard deviation $\sigma$. Given a dataset of $r+1$ estimators, the order of the remainder $R$ in Eq.~\eqref{eq:remainder-term-theorem} would be $2(r+1)$ corresponding to the $2r+1^\mathrm{th}$ degree Lagrange interpolating polynomial with respect to $\sigma$ and the odd-order terms set to zero.

We now apply this noise model to the time evolution under a many-body spin Hamiltonian in a chain of Rydberg atoms. In such experiments, atoms are regularly spaced in a one-dimensional configuration using optical tweezers or lattices. A spin model is then realized by coupling the ground electronic state $\ket{\downarrow}$ to a high-lying Rydberg state $\ket{\uparrow}$ using lasers that globally address all atoms to drive one- or two-photon transitions at Rabi frequency $\Omega$. Strong two-body van der Waals interactions prevent nearby atoms from being simultaneously excited to the Rydberg state. Due to this blockade effect, in the limit where only nearest-neighbor atoms are blockaded and neglecting small beyond-nearest-neighbor interactions, the system Hamiltonian may be written~\cite{bernien_probing_2017} as
\begin{equation}
    \hat H=\Omega \left(\hat \sigma^x_1\hat n_2 +\sum_{i=2}^{L-1} \hat n_{i-1}\hat\sigma^x_{i}\hat n_{i+1}+\hat n_{L-1}\hat\sigma^x_L\right),
\end{equation}
where $\hat n_i=\frac{1}{2}\left( \hat{\mathbb{1}}-\hat \sigma^z_i\right)=\ket{\downarrow}\bra{\downarrow}_i$ is the projector on the ground state of the $i^\mathrm{th}$ atom. An interesting dynamical effect in this model occurs when starting from the Néel state $\ket{\downarrow \uparrow \cdots \downarrow\uparrow}$: although this is an infinite-temperature state (i.e.~it has largest overlap with states in the middle of the spectrum), and therefore is expected to thermalize rapidly~\cite{reimann2016}, it exhibits long-lived oscillations of the staggered magnetization
\begin{equation}    S^\mathrm{z}_\mathrm{stag}=\sum_{i=1}^L (-1)^i \braket{\hat\sigma^z_i}.
\label{eq:stagmag}
\end{equation}
These long-lived oscillations result from a large overlap of the Néel state with a set of special high-energy eigenstates known as many-body scars that lead to slow thermalization~\cite{turner2018}. The frequency of these oscillations is proportional to $\Omega$, as this is the only energy scale in the problem. Therefore, shot-to-shot fluctuations in $\Omega$ can directly lead to a noise-induced excess decay of the many-body-scar oscillations, analogous to the single-qubit case in Fig.~\ref{fig:1}(a).

Experimentally, the Rabi frequency $\Omega$ in Rydberg atom experiments is proportional to the electric field amplitude of the laser beam. Hence, laser intensity fluctuations on timescales comparable to the repetition rate of the experiment lead to shot-to-shot variations of $\Omega$ as illustrated in Fig.~\ref{fig:1}(c). In recent Rydberg-atom experiments, it was shown that such fluctuations are the leading error source in such experiments, and these are usually Gaussian distributed~\cite{shaw2024,wurtz2023}. 

To show the influence of shot-to-shot noise on many-body-scar dynamics, we numerically calculate the staggered magnetization as a function of time, for various values of $\sigma^2$, shown in \begin{figure}[t] 
    \centering
\includegraphics{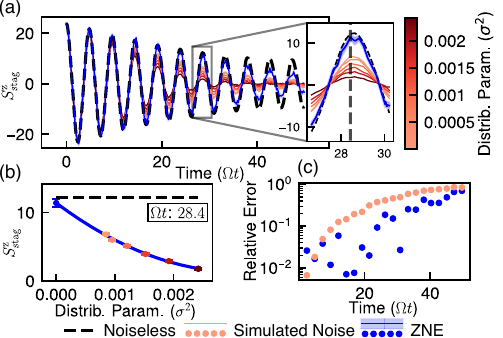}
    \caption{\textbf{ZNE of many-body scars in Rydberg atoms.} Numerical data from exact diagonalization, system size $L=24$. (a) Staggered magnetization $S^z_\mathrm{stag}$ [Eq.~\eqref{eq:stagmag}] in the model for normally distributed shot-to-shot noise (shades of red; curve  with smallest noise shown in legend and has standard deviation $\sigma_0= 3\%$), noiseless oscillations (black dashed), and the result of zero-noise extrapolation (blue) with shaded error bars given by the error of the polynomial fit (light blue). (b) Staggered magnetization plotted against the noise strength (distribution parameter) $\theta=\sigma^2$ for a single time $t$ such that $\Omega t=28.4$ and fit to a polynomial via leave-one-out cross validation (LOOCV, blue line). Error bars are simulated quantum projection noise with $N=1000$ (shades of red)  and error of the polynomial fit (blue). (c) Relative error at the maxima and minima of the oscillations for the ZNE results and the result with the smallest noise, $\sigma_0=3\%$.}
    \label{fig:2}
\end{figure}
Fig.~\ref{fig:2}(a). We find that, as $\sigma$ increases, the decay rate of oscillations increases: after applying ZNE, the lifetime of the observed oscillations is extended to longer times $\Omega t$. 

In order to test the performance of the analog ZNE in this system, we show the staggered magnetization as a function of noise strength for a fixed time in Fig.~\ref{fig:2}(b). We find a relatively smooth behavior even in the presence of a finite number of shots $N=1000$. Fitting a polynomial, we extrapolate to zero noise, finding a zero-noise value that is closer to the noiseless result than the noisy result with the lowest $\sigma^2$. We assume $\sigma_0=3\%$ for the smallest experimentally realizable standard deviation. This is realistic as standard deviations as small as $0.5\%$ have been measured~\cite{shaw2024, wurtz2023}. Repeating this for all time points, we find a good match with the noiseless results up to long times (see blue line in Fig.~\ref{fig:2}(a). We also plot the relative error as a function of time in Fig.~\ref{fig:2}(c), finding that the increase of the error is substantially slowed with zero-noise extrapolation compared to the noisy result. In particular, we find that the time at which the error in the staggered magnetization reaches $10\%$ is increased by a factor of three. 

In practice, this scheme requires (1) an accurate $x$-axis for extrapolation, and (2) controllably enhancing the noise. For (1), while the theoretical framework permits extrapolation based on scaling the noise parameter $\theta$ relative to a potentially unknown baseline $\theta_0$ (defining a dimensionless axis $c$ such that $c_i \theta_0 = \theta_i$), experimental realization of this relative scaling is nontrivial. We circumvent this by directly learning $\theta$ through measurements of single-qubit Rabi oscillations between the ground and Rydberg state and fitting to Eq.~\eqref{eq:Gauss}. While this makes use of analytic results for single qubit evolutions which may not exist for multi-qubit evolutions, $\theta$ is system size independent and remains a valid extrapolation parameter for general multi-qubit perturbations. For (2), one can utilize a feedback controller of the Rydberg laser in order to control the spectral feature of low-frequency noise. A more direct approach is to slightly change the intensity of the laser in each shot of the experiment, with the change distributed according to a Gaussian. This artificial shot-to-shot fluctuation adds to the naturally occurring Gaussian shot-to-shot noise. Because the probability distribution of the sum of two random, Gaussian-distributed variables is also Gaussian distributed, this enables enhancing the noise as required. As an example, the QuEra Aquila~\cite{wurtz2023} laser control is sufficient to implement this simple scheme.

\subsection{Thermal noise in individually-addressed trapped ions}\label{sct:ions}

We now apply our analog zero-noise extrapolation to experiments on trapped ions with individual ion addressing. If the ions are not cooled to their motional ground state, they move within the laser beam, leading to a fluctuating laser intensity on each ion~\cite{Seetharam2024,cetina2022}, c.f. Fig.~\ref{fig:1}(c). For quantum simulators that use large Raman beams addressing the chain of ions globally, the programmability of interaction is limited,  and this noise is small. For advanced simulators, however, ion addressing and control relies on arrays of tightly focused laser beams, for which fluctuations  in the atom-light coupling strength are noticeable. In long chains, this becomes the dominant source of error \cite{cetina2022}. The initial ion motional state, which fluctuates for each experimental shot, determines the effective Rabi frequency sensed by an ion; the average of these fluctuations over the thermal distribution leads to the simulation error.

To quantitatively describe this noise model, we summarize the framework presented in Ref.~\cite{cetina2022} which describes the Rabi frequency errors of ions in tightly focused beams due to fluctuating motion along the chain axis. The ions' motion often results from imperfect cooling (state preparation of the motion in transverse directions) and from anomalous heating~\cite{Deslauriers2006, McConnell2015, Boldin2018, brownnutt2015}. The latter process, associated with noisy background electric fields that drive the ions at their motional resonance frequencies, acts predominantly on the axial center-of-mass (COM) phonon mode which has the longest wavelength and the lowest frequency. Consequently, the axial COM motional mode retains a high average phonon occupation $\bar{n}$ of hundreds of quanta for moderate chains, scaling sharply with the size of the ion chain; by contrast the radial modes are prepared via sideband cooling to below one quantum on average. Assuming that the axial COM dominates the fluctuating ion motion, we cast the local Rabi frequency of the $i^\mathrm{th}$ ion as
\begin{equation}
    \tilde{\Omega}_i(n) =\Omega_{i} + \frac{1}{2}\Omega''_{i}b_{i}^2\frac{\hbar(n + 1/2)}{M\omega}.\label{eq:fock-state-omega}
\end{equation}
The first term, $\Omega_{i}$, represents the ideal Rabi frequency in the absence of ion motion. The second term represents corrections due to the motion of ions within the beam. $\omega$ denotes the angular frequency of the axial COM mode; $\Omega_{i}''$ is the spatial second derivative of the beam profile at the ion's average position; $b_{i}$ is the mode participation factor of ion $i$ to the COM mode as defined in Ref.~\cite{monroe2021}, and $M$ is the mass of a single ion. Here $n$ is the number of phonons in the COM mode at a given experimental shot with a thermal average $\bar{n}$. Fluctuation of $n$ between experimental shots leads to shot-to-shot noise.

To understand the impact of the shot-to-shot fluctuations on Rabi oscillations, we now evaluate the averaged transfer probability in analogy with Eq.~\eqref{eq:Gauss}. Averaging the Rabi oscillation over the thermal distribution $p(n)\propto e^{- \hbar\beta\omega n}$ ($\beta = 1/(k_\mathrm{B}T)$ is the inverse temperature and $k_\mathrm{B}$ is the Boltzmann constant), in the limit of high temperatures such that $\bar n\approx 1/(\hbar \beta\omega)$, Ref.~\cite{cetina2022} derived an analytic expression:
\begin{equation}\bar P_{\uparrow}(t) =  \frac{1}{2}\left(1-C(t,\theta)\cos\left(\Omega_{i}t + \phi(t,\theta)\right)\right)\label{eq:thermal_rabi}
\end{equation}
with $C(t,\theta) = \frac{1}{\sqrt{1 + (\Omega_{i}\theta t)^2}}$, $\phi(t,\theta) = \arctan(\Omega_{i}\theta t)$, and 
\begin{equation}
    \theta = \alpha \bar{n}.
    \label{eq:thetadefthermal}
\end{equation}
Here, $\alpha = -b_{i}^2 \hbar \Omega''_{i}/\left(2M\omega\Omega_{i}\right)$ is a system dependent scale parameter $|\alpha| \ll 1$ that describes the linear relationship between $\bar n$ and perturbations to $\Omega$. From this expression, we see that $\theta$ plays a similar role to the variance in the Gaussian distribution case, Eq.~\eqref{eq:Gauss}, as it determines the decay rate of the envelope of the oscillations. Therefore, we use $\theta$ as our distribution parameter.

Despite the non-Gaussian nature of the noise, we obtain the following corollary of Theorem~\ref{thm:main}---proven in Appendix~\ref{app:proofs}---demonstrating that the thermal distribution is mitigable and that ZNE does apply to this form of noise, defined for the Hamiltonian $\hat H = \hat H_0 + n \hat V$, with integers $n$ drawn from the thermal distribution with probability mass function
\begin{equation}
    \mathrm{Pr}_{\bar n}(n)=\frac{1}{\bar{n}+1}\left( \frac{\bar{n}}{\bar{n}+1} \right)^{n} \label{eq:Prdef}.
\end{equation}

\begin{tcolorbox}[colback=Maroon!10, colframe=gray!100,rounded corners=all,boxrule=0.5mm, width=\columnwidth]
\begin{corollary}[ZNE for thermally distributed noise]
    Thermal noise is mitigable for times $t < \left(2(\bar{n}+1)\|\hat{V}\|_\infty\right)^{-1}$ and permits zero-noise extrapolation along the $\theta =\bar{n}$ axis with remainder \begin{equation}
        \left|R_{r+1}(\bar{n}, t, \hat V)\right| \leq \|\hat{O}\|_\infty\left(2(\bar{n}+1)t\|\hat{V}\|_\infty \right)^{r+1}.
    \end{equation} \label{cor:thermal-ZNE}
\end{corollary}
\end{tcolorbox}
In contrast to Cor.~\ref{cor:ZNE-Normal} and Eq.~\eqref{eq:defshot}, where it is implied that $\delta \ll 1$ and the extrapolation parameter is the variance $\sigma^2 \ll 1$, for thermal noise the extrapolation parameter $\bar n$ is typically large. This in turn requires $\|\hat V\|_\infty$ be small. Indeed, for the case of the noise model for surface electrode trapped-ion systems captured in Eq.~\eqref{eq:thermal_rabi}, $\|\hat V\|_\infty \propto \alpha \ll 1$. However, the same equation suggests the true experimentally relevant parameter is the rescaled value $\theta = \alpha \bar n$, which, as we will demonstrate, can be directly measured from experimental results. 

Furthermore, the result for thermally distributed noise provides a finite lower bound on the radius of convergence of $\mathrm{Tr}(\hat O\hat \rho_\theta(t))$ around $\theta=0$. This is a direct result of the $k^\mathrm{th}$ thermal distribution moment scaling as $k!\theta^k$, which offsets the factorial term in Eq.~\eqref{eq:main-theorem-statement}. Consequently, thermal ZNE, unlike ZNE for Gaussian shot-to-shot noise, cannot achieve arbitrarily strong noise mitigation in the limit of zero projection noise, $\nu = 0$, and $r \to \infty$. 

The thermal distribution additionally has a non-zero average value $\theta$ which contributes to the noisy time-dependent phase accumulation, $\phi(t, \theta)$, in Eq.~\eqref{eq:thermal_rabi}. Although accounting for $\theta$ is not required for Theorem~\ref{thm:main}, it contributes an error with a constant frequency component that may be corrected before performing the ZNE in order to reduce the remainder $R$ by a constant factor converging to $e$ for large values of $r$ (see Appendix~\ref{app:rabi-correction}). In the experimental implementation of our method, we account for this correction by performing ZNE as a two-step process: first, we calibrate the experimental Rabi frequency $\Omega_i$ in a $\theta$-dependent way such that the frequency of the Rabi oscillation is approximately independent of $\theta$. Then, we perform ZNE on the $\bar n$ dependence of $C$. We demonstrate this procedure experimentally for both single-qubit Rabi oscillations and two-qubit exchange oscillations.

\subsubsection{Extracting \texorpdfstring{$\theta$}{theta} and mitigating single-qubit Rabi oscillations}

We first study Rabi oscillations, equivalent to the evolution of spins in a magnetic field, in the presence of thermal noise. We use a chain of 27 ions trapped above a surface trap and illuminated with an array of optical beams perpendicular to the chain axis, which drive spin transitions. Our experimental setup has been described previously in Refs.~\cite{Feng2023,katz2024observing, schuckert2025} with additional details in Appendix~\ref{app:experiment}. 

Single-qubit Rabi oscillation measurements are presented in 
\begin{figure*}[t]
    \centering
\includegraphics[width=2\columnwidth]{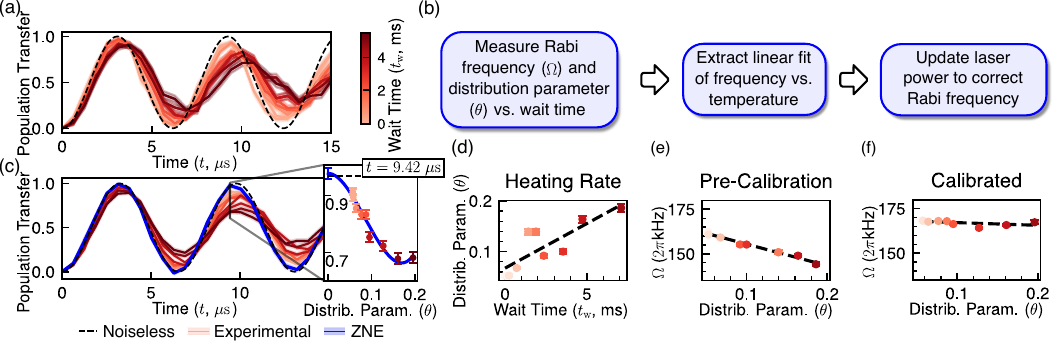}
    \caption{\textbf{Experimental ZNE of Rabi oscillations for thermal noise in trapped ions.} Data reported from a single ion in a $27$-ion linear chain. The central $23$ ions are driven simultaneously.
   (a) Single-qubit Rabi oscillations for different wait times (increasing color shade), compared to exact expectation (black dashed line). (b) Procedure for calibrating Rabi frequency as a function of the distribution parameter $\theta$. From calibration measurements over a range of wait times, the Rabi frequency $\Omega$ and the distribution parameter $\theta$ are extracted. The input laser amplitude is then adjusted for each wait time to enable the first-order frequency correction. (c) Full Rabi oscillations obtained with the wait-time-dependent Rabi frequencies obtained through the procedure in (b). Blue line is the ZNE result, obtained with polynomial order determined by LOOCV and the linear component set to zero. Inset: Third-order polynomial fit for $t=9.4~\mu$s. (d) From the data in (a), the noise distribution parameter $\theta$ as a function of initial wait time and the resulting linear fit. (e) Also from the data in (a), Rabi frequencies plotted against the distribution parameter. (f) From the data in (c), Rabi frequencies after the first-order correction plotted against the distribution parameter. 
    \label{fig:3}}
\end{figure*}
Fig.~\ref{fig:3}(a). To demonstrate the temperature dependence of the decay rate, we controllably heat the COM mode by introducing a wait time $t_\mathrm{w}$ between cooling the ions' motional modes and the onset of the Rabi oscillation. During the delay, the Rabi drive is turned off. As expected, we find that the decay rate of the oscillations increases as the wait time is increased. By fitting the experimental results to Eq.~\eqref{eq:thermal_rabi} we extract $\theta$ and $\Omega_{i}$ as a function of the wait time, Fig.~\ref{fig:3}(d,e). We extract a linear fit $\theta = \dot{\theta}t_\mathrm{w} + \theta_0$ by measuring the value of $\theta$ for a range of experimental waiting times $t_\mathrm{w}$. The extrapolation process corresponds to the limit $\theta \to 0$. Crucially, varying $t_{\mathrm{w}}$ provides a means of specifically manipulating $\theta$ while leaving other Hamiltonian or error terms invariant. Additionally, the heating only happens on a timescale of $\sim 1$~ms, which is much slower than that of Rabi oscillations. 

Having extracted $\theta$, we now describe how to compensate for the time-dependent phase $\phi(t)$, Eq.~\eqref{eq:thermal_rabi}. We do so by approximating its effect as a shift of the Rabi frequency, which provides a good approximation for short times. We compensate for this effect by simply calibrating the Rabi frequency for each waiting time individually. We show the pre- and post-calibration oscillation frequencies in Fig.~\ref{fig:3}(e) and (f), respectively. In a multi-qubit experiment, this procedure generalizes to the case with two-body interactions and is done for each ion individually due to the dependence of $\theta$ on the ion index through the mode participation factors $b_i$, see Appendix~\ref{app:rabi-correction}. This procedure effectively centers the thermal distribution, similar to the process described in Ref.~\cite{Seetharam2024} for digital circuits. Afterwards, we apply our ZNE procedure to extrapolate to the limit of zero variance. 

It is important to note that while the derivation of $\theta$ and our initial compensation procedure here rely on single-qubit analytic results, the underlying principle of ZNE for this noise is broadly applicable. The ability to identify and controllably vary a dominant noise parameter, in this case $\theta$ via $t_\mathrm{w}$, is key. This allows us to apply the same ZNE framework to many-body interacting Hamiltonians for which no analytic solutions are known. Therefore, even for these single-qubit demonstrations, we intentionally perform polynomial extrapolations rather than fitting directly to the analytic form of Eq.~\eqref{eq:thermal_rabi}, underscoring the generality of the method.

We apply ZNE to the Rabi-frequency-corrected results in Fig.~\ref{fig:3}(c, f). We find almost perfect agreement between the ZNE and the noiseless results, showing that the error from thermal noise can be efficiently reduced. 

\subsubsection{Mitigating effects from thermal motion in two-qubit interactions}\label{sct:exp_twoq}

We now apply our analog ZNE to a two-qubit problem in trapped ions.
By detuning the Raman beams, we drive virtual-phonon-mediated interactions between the ion qubits. This interaction is given by the Hamiltonian
\begin{equation}
    \hat{H} \propto \Omega_{0}\Omega_{1}\hat\sigma^x_0\hat \sigma^x_1 \label{eq:two-qubit-ham},
\end{equation}
i.e.~it is proportional to the Rabi frequencies of the two ions. Due to its low frequency, the heating in the COM mode dominates the phonon heating. As such, the perturbations to the Rabi frequency are highly spatially correlated and the error in $\Omega_0\Omega_1$ for two separate ions indexed $0$ and $1$ is well approximated by a single perturbation $\Omega_0\Omega_1 \to \Omega_0\Omega_1(1 + \delta)$, with $\delta$ drawn from a thermal distribution. 

Our goal is to mitigate these fluctuations. In order to do so, we prepare the initial state $\ket{\uparrow \downarrow}$ and measure the observable $\hat Z _{\text{diff}} \equiv (\hat\sigma^z_0 - \hat\sigma^z_1)/2$. In addition to shot-to-shot errors, the system is also subject to slow incoherent bit-flip errors. To assess the performance of the ZNE independently of these bit-flip errors, we perform post-selection on the $\{\ket{\uparrow \downarrow}, \ket{\downarrow \uparrow}\}$ subspace to remove single bit-flip errors. Assuming only single bit-flip errors happen, this post-selection does not introduce additional error into the ZNE beyond an increase in the variance of the individual estimators. 

We show the time-evolution of $\langle \hat Z _{\text{diff}} (t) \rangle$ in 
\begin{figure}[t]
    \centering
\includegraphics[width=\columnwidth]{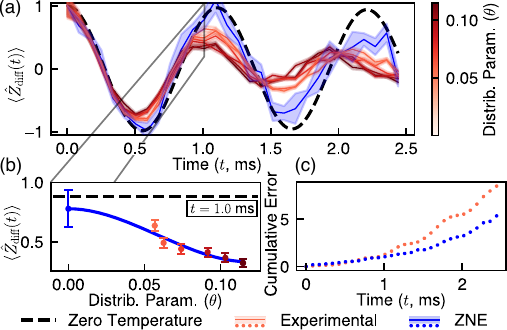}
    \caption{\textbf{Experimental zero-noise extrapolation for two-qubit Ising interactions in trapped ions.} (a) Magnetization difference between the two qubits as a function of time. To mitigate bit-flip errors, we use post-selection onto states $\ket{\downarrow \uparrow}$ and $\ket{\uparrow \downarrow}$, keeping $380-460$ out of $600$ shots. We calibrate the Rabi frequency using the single-qubit Rabi oscillations as shown in Fig.~\ref{fig:3}. 
    Zero-noise extrapolation result using LOOCV (first-order polynomial component is set to zero) is given by the blue curve, with error bars given by the standard deviation of the least-squares polynomial fit. The black curve is a simulation without shot-to-shot noise but with a bit-flip error matched to the observed experimental post-selection rate and subsequent numerical post-selection. (b) The result of the extrapolation process at time $t=1$ ms, which is a third-order fit with the first-order contribution set to zero. (c) Cumulative error at time $T$ defined as $\sum_{t=0}^T|\langle \hat Z ^\mathrm{exact}_{\text{diff}} (t) \rangle-\langle \hat Z _{\text{diff}} (t) \rangle|$, where $\langle \hat Z ^\mathrm{exact}_{\text{diff}} (t) \rangle$ ($\langle \hat Z _{\text{diff}} (t) \rangle$) denotes the exact (noisy) value of the measured observable at time $t$. The orange dots correspond to $\theta_0\approx0.05$, and the blue dots to the result of the ZNE.}
    \label{fig:4}
\end{figure}
Fig.~\ref{fig:4}(a-c). We find a strong damping of the oscillations which we attribute to shot-to-shot fluctuations of the Rabi frequencies. Indeed, by increasing the wait-time (equivalently, temperature or distribution parameter) as in Fig.~\ref{fig:3}, we find a stronger decay of the oscillations as expected. We then apply our analog zero-noise extrapolation to such two-qubit oscillations with post-selection. In order to approximately correct the first order frequency error as a function of wait time, we further apply the Rabi-frequency correction described in Fig.~\ref{fig:3} for each wait-time dataset before the two-qubit measurements are performed. The results of the ZNE demonstrate partial recovery of the contrast in the two-qubit oscillations, with reduced error compared to the lowest temperature (baseline) measurement. For early times, the extrapolation under-performs compared to the baseline results. This is due to the fact that projection noise dominates in this regime. However, ZNE significantly improves quality of the result near the first full period, where the contrast has substantially decayed without ZNE. For longer evolution times, we still find an improvement of the contrast, but the contrast is less well recovered than the single-qubit Rabi oscillations in Fig.~\ref{fig:3}. We attribute this to the substantially longer time scale of this experiment ($1$ ms vs. $10~\mu$s), highlighting the role of other error effects, such as dynamic heating within each shot or Markovian errors, playing a larger role here.

\section{Discussion and Outlook \label{sec:discussion}}
In this work, we introduced analog zero-noise extrapolation (ZNE) as a framework for mitigating shot-to-shot errors with quasi-static temporal correlations in analog quantum simulators. We proved that observables can be expanded in a polynomial series of a suitably defined shot-to-shot noise distribution parameter for sufficiently small fluctuations. This proof shows that ZNE can be performed by amplifying the distribution of the noise and extrapolating measured observables to the zero-noise limit. Numerically simulating ZNE for shot-to-shot noise in intensity of the laser driving transitions to a Rydberg state, we predict an extended lifetime of many-body scar oscillations by a factor of three in Rydberg atom arrays. We implement analog ZNE in a trapped-ion experiment where thermal motion induces shot-to-shot fluctuations of single-qubit Rabi frequencies and inter-ion interactions, finding a large enhancement of the effective coherence. These results demonstrate an experimentally feasible protocol to mitigate the effects of a key source of noise limiting the scalability of many analog simulation platforms. For example, in individually-addressed trapped-ion systems, thermal motion induces shot-to-shot error in two-body entangling interactions scaling as $N^6$ with the number of ions $N$~\cite{cetina2022}.
In the absence of a systematic approach for scalable error correction in analog quantum dynamics, such error mitigation schemes will always be applicable to cutting-edge experimental systems, and mechanisms to address leading error sources will be valuable in extending the effective lifetime of target dynamics. This approach can help push analog quantum simulations into regimes currently inaccessible to classical methods, especially for systems like Rydberg atoms and long chains of individually-addressed trapped ions where the shot-to-shot noise discussed in this work is a dominant error source.

Looking ahead, multiple avenues are opened up. It is beneficial to implement analog ZNE in a large-scale quantum simulation experiment limited by such fluctuations. This method may also be integrated with  multi-parameter ZNE protocols~\cite{russo_quantum_2024} to address multiple noise sources simultaneously, in particular incoherent Markovian noise alongside shot-to-shot fluctuations; also, laser frequency fluctuations in neutral-atom arrays, manifesting as a global shift $\Delta \sum_i \hat{\sigma}^z_i$, could be mitigated at the same time as intensity fluctuations. Beyond this, it is known that thermal motion of neutral atoms and molecules in optical tweezers can lead to independent variations in the interaction rate between different sites~\cite{holland_-demand_2023}, while tweezer intensity fluctuations may result in site-dependent chemical potential terms relevant for simulations of models with lattice hopping terms~\cite{murmann_two_2015,spar_realization_2022}. These latter cases are not directly covered by our proof but we expect it to generalize straightforwardly. Shot-to-shot noise can also occur in condensed-matter contexts, for example in THz-driven donors~\cite{Greenland2010, crane2021}, where amplification of the noise is less straightforward. Lastly, there are physically relevant noise distributions that go beyond our framework due to the scaling of their moments. For example, the Cauchy-Lorentz distribution does not have well defined moments and so is not captured by Theorem~\ref{thm:main}; yet due to its connection to Markovian dynamics standard ZNE methods may be sufficient. 

\section{Acknowledgments}

We acknowledge helpful discussions with Timothy Connor Mooney, Matthew Diaz, Alec Cao, Marko Cetina, and Pedro Lopes.
This material is based upon work supported by the U.S.~Department of Energy, Office of Science, National Quantum Information Science Research Centers, Quantum Systems Accelerator (QSA). 
Additional support is acknowledged from a seed grant funded through NSF OMA-2120757 (Quantum Leap Challenge Institute for Robust Quantum Simulation). T.S.~and A.V.G.~were also supported in part by the DoE ASCR Quantum Testbed Pathfinder program (awards No.~DE-SC0019040 and No.~DE-SC0024220), NSF QLCI (award No.~OMA-2120757), NSF STAQ program, AFOSR MURI, ARL (W911NF-24-2-0107),  DARPA SAVaNT ADVENT, and NQVL:QSTD:Pilot:FTL. T.S.~and A.V.G.~also acknowledge support from the U.S.~Department of Energy, Office of Science, Accelerated Research in Quantum Computing, Fundamental Algorithmic Research toward Quantum Utility (FAR-Qu). Y.-X.W.~acknowledges support from a QuICS Hartree Postdoctoral Fellowship. S.R.M. is supported by the NSF QLCI (award No. OMA-2120757). Specific product citations are for the purpose of clarification only, and are not an endorsement by the authors or NIST.
\textbf{Data availability.} All data and codes are available on reasonable request.
\bibliography{bib}
\FloatBarrier

\appendix
\setcounter{theorem}{0}
\setcounter{lemma}{0}
\setcounter{corollary}{0}
\begin{widetext}
\section{Order finding for zero-noise extrapolation}\label{app:LOOCV}

Throughout this work, we make use of least-squared regression for fitting extrapolation polynomials. We observe that Richardson extrapolation with projection noise is prone to over-fitting while least-squares regression performs well and is more stable, typically using polynomials of degree much smaller than used in Richardson extrapolation. The specific degree of the polynomial used for extrapolation in each dataset was determined using an adapted leave-one-out cross validation method (LOOCV) intended to both reduce (human) bias in determining the appropriate polynomial order and to reduce overfitting. 

The method used for LOOCV is as follows. For each time $T_k$ and dataset $\{(\theta_i, \tilde{E}_{\theta_i,T_k})\}_{i=0}^r$, and for each polynomial degree $0\leq p_m < r$, we leave out a single point indexed $j$, one at a time for all $0\leq j\leq r$ and fit a polynomial of degree $p_m$, which we call $\mathrm{polyfit}(p_m,{\{(\theta_i, \tilde{E}_{\theta_i}(T))\}_{i\neq j}^r})$, and compute the residual in estimating the left-out point:
\begin{equation}
    \mathrm{Res}(m) = \sum_{j=0}^r \left(\frac{\tilde{E}_{\theta_j} -\mathrm{polyfit}(p_m,\{(\theta_i, \tilde{E}_{\theta_i}(t))\}_{i\neq j}^r}{\nu_j}\right)^2, 
\end{equation}
where $\nu_j$ is the standard error 
\begin{align}
 \nu_j^2 = \frac{   \mathrm{Tr(\hat{O}^{2}\rho_{\theta_{i}}(t))}-\mathrm{Tr(\hat{O}\rho_{\theta_{i}}(t))^{2}} }{N_j}.
\end{align}The polynomial degree $p_{k}$ which minimizes the residual is then chosen for each time point $t_k$. Due to the dependence of the error on the simulation time $t_k$, we perform a linear fit $at_K +b = p'_k $ of $t_k$ against $p_k$, so the degree of the fit is monotonically increasing with time, and use $\lceil p'_k\rceil$ as the final order used in the ZNE. \label{appendix: LOOCV}

\section{Proof of Theorem 1 and its corollaries}
\label{app:proofs}
In this appendix, we prove our main theorem as well as its corollaries.

\subsection{Proof of Theorem~\ref{thm:main}}
\begin{theorem}
    Consider a set of $r+1$ estimators (each averaged over a large number of shots) on an analog quantum simulator subject to shot-to-shot noise drawn from a mitigable distribution $\mathcal{D}_{\theta}$, resulting in pairs of noise strengths and noisy estimators, $\{(\theta_{0}, \tilde{E}_{\theta_{0},t}), \dots, (\theta_{r}, \tilde{E}_{\theta_{r},t})\}$, ordered by the value of $\theta_i$, each with sampling error distributed with variance $\nu_{i}^2$. Then, with probability $1-\epsilon$, the result of ZNE, i.e.~the Richardson extrapolator and absent additional information about $\mathcal{D}_\theta$, has an error bounded by:
\begin{align}
    |\tilde{E}_{0,t}-\mathrm{Tr}&(\hat{O}\hat\rho_0(t))| \nonumber \\ &\leq \sum_{i=0}^{r}|\gamma_{i}|\left( |R_{r+1}( \mathcal{D}_{\theta_i},t,\hat{V})| + c_i \right),
\end{align}
where the remainder error term $R$ is upper bounded by the absolute value of raw moments of the distribution $\mathcal{D}_{\theta},$
\begin{align}
    |R_{r+1}(\mathcal{D}_{\theta_i}, t, \hat{V})| &\leq \|\hat{O}\|_{\infty}\frac{(2t)^{r+1}\|\hat{V}\|_{\infty}^{r+1}\mathbb{E}_{\theta_i}\left[ |\delta^{r+1}|\right]}{(r+1)!},
\end{align}
and $c_i = \sqrt{ 2 }\nu_i \mathrm{erf}^{-1}(1-\epsilon)$ bounds the quantum sampling error in the large-shot-number limit. $\|\hat A \|_\infty$ is the spectral norm equal to the largest eigenvalue norm of $\hat A$.
\end{theorem}

\noindent\textbf{Proof:}
We employ a strategy related to zero-noise extrapolation for noise from a super-operator $\mathcal{L}$ presented in Ref.~\cite{temme2017} and extend the results to the case of sampling from shot-to-shot noise. 

In order to consider the influence of the error perturbatively in the factor $\delta$, we first move into the interaction picture with respect to $\hat H_0$ by defining $\hat U_0=\exp(-i\hat H_0 t)$ and
\begin{align}
    \rho_\mathrm{I}(t) &= U_{0}^{\dagger}(t)\rho(t)U_{0}(t), \\
    \hat{A}_\mathrm{I}(t) &= U_{0}^{\dagger}(t)\hat{A}U_{0}(t).
\end{align}
Note that $\mathrm{Tr}(\hat{A}_\mathrm{I}\rho_\mathrm{I}(t))= \mathrm{Tr}\left( \hat{A}\rho(t) \right)$. Time evolution in the interaction picture is given by 
\begin{equation}
    \rho_\mathrm{I}(t) = \rho_\mathrm{I}(0) -i \delta \int _{0}^{t} \mathrm{d}t_1 [\hat{V}_\mathrm{I}(t_1), \rho_\mathrm{I}(t_1)]\, . 
\end{equation}
We now expand the time evolution for small $\delta$. Specifically, we write the Dyson series up to $r^\mathrm{th}$ order:
\begin{align}
    \rho_\mathrm{I} (t) = \rho_\mathrm{I} (0) +& \sum_{k=1}^{r} (-i\delta)^k \int_0^t\dots\int_0^{t_{k-1}} \mathrm{d}t_k \dots \mathrm{d}t_1  [\hat{V}_\mathrm{I}(t_1),[ \dots , [\hat{V}_\mathrm{I}(t_{k}), \rho_\mathrm{I}(0)]\dots]] \notag\\
    +& (-i\delta)^{r+1}\int_0^t \dots \int_0^{t_{r}} \mathrm{d}t_{r+1}\dots \mathrm{d}t_1[\hat{V}_\mathrm{I}(t_1),[ \dots , [\hat{V}_\mathrm{I}(t_{r+1}), \rho_\mathrm{I}(t_{r+1})]\dots]].\label{eq:remainder_term_initial}
\end{align}
The last term is the remainder of the expansion, which we aim to rigorously bound in the following discussion. 
Given the initial density matrix $\rho_0 = \rho_I(0)$, we now measure the interaction picture operator $\hat{O}_{I}(t)$. We can then make use of the cyclic property of the trace to formulate the noisy and noise-free expectations in the Schrödinger picture:
\begin{align}
    \mathrm{Tr}\left(\hat{O}\rho(t)\right) = &\mathrm{Tr}\left(\hat{O}\rho_{\delta=0}(t)\right)+  \sum_{k=1}^{r} (-i\delta)^k  \int_0^t\dots\int_0^{t_{k-1}} \mathrm{d}t_k \dots \mathrm{d}t_1 \mathrm{Tr} \left( \hat{O}_I(t) [\hat{V}_\mathrm{I}(t_1), \dots , \hat{V}_\mathrm{I}(t_k), \rho_0]\right) \notag\\
    &\qquad+ (-i\delta)^{r+1}\int_0^t \dots \int_0^{t_{r}} \mathrm{d}t_{r+1}\dots \mathrm{d}t_1 \mathrm{Tr}\left(\hat{O}_I(t)[\hat{V}_\mathrm{I}(t_1),[ \dots , [\hat{V}_\mathrm{I}(t_{r+1}), \rho_\mathrm{I}(t_{r+1})]\dots]]\right),
\end{align}
which notably includes measurements on the noisy observable $\mathrm{Tr}\left(\hat{O}\rho(t)\right)$ and the noise-free observable $\mathrm{Tr}\left(\hat{O}\rho_{\delta=0}(t)\right)$. Next, we average over the noise distribution $\mathcal{D}_{\theta}$:
\begin{align}
    \mathbb{E}_{\theta}\left[ \mathrm{Tr}\left(\hat{O}\rho(t)\right)\right] &= \mathrm{Tr}\left(\hat{O}\rho_{\theta=0}(t)\right)\notag\\&\quad+  \sum_{k=1}^{r} \mathbb{E}_{\theta}[\delta^k](-i)^k  \int_0^t\ \dots\int_0^{t_{k-1}} \mathrm{d}t_k \dots \mathrm{d}t_1  \mathrm{Tr} \left( \hat{O}_{\mathrm{I}}(t) [\hat{V}_\mathrm{I}(t_1),[ \dots , [\hat{V}_\mathrm{I}(t_{k}), \rho_\mathrm{I}(0)]\dots]]\right) \notag\\
    &\quad+ \mathbb{E}_{\theta}\left[\delta^{r+1}(-i)^{r+1}\int_0^t \dots \int_0^{t_{r}}\mathrm{d}t_{r+1}\dots \mathrm{d}t_1 \mathrm{Tr}\left(\hat{O}_{\mathrm{I}}(t) [\hat{V}_\mathrm{I}(t_1),[ \dots , [\hat{V}_\mathrm{I}(t_{r+1}), \rho_\mathrm{I}(t_{r+1})]\dots]]\right)\right].
\end{align}
Each term in the Dyson series expansion can be written as a power series in the moments of the distribution $\mathcal{D}_{\theta}$ by defining 
\begin{align}
a_k =(-i)^k  \int_0^t\dots\int_0^{t_{k-1}}  \mathrm{d}t_k \dots \mathrm{d}t_1 \mathrm{Tr}\left(\hat{O}_{\mathrm{I}}(t) [\hat{V}_\mathrm{I}(t_1),[ \dots , [\hat{V}_\mathrm{I}(t_{k}), \rho_\mathrm{I}(0)]\dots]]\right) 
\end{align}
to give
\begin{align}
   \mathbb{E}_{\theta} \left[\mathrm{Tr}\left(\hat{O}\rho(t)\right)\right] 
    &=
    \mathrm{Tr}\left(\hat{O}\rho_{\theta=0}(t)\right)+  \sum_{k=1}^{r} \mathbb{E}_{\theta}[\delta^k] a_k \notag\\
    &\quad+ 
    \mathbb{E}_{\theta}\left[
        \delta^{r+1}(-i)^{r+1}\int_0^t \dots \int_0^{t_{r}}\mathrm{d}t_{r+1}\dots \mathrm{d}t_1 \mathrm{Tr}\left(\hat{O}_{\mathrm{I}}(t) [\hat{V}_\mathrm{I}(t_1),[ \dots , [\hat{V}_\mathrm{I}(t_{r+1}), \rho_\mathrm{I}(t_{r+1})]\dots]]\right)\right].\label{eq:series_expansion}
\end{align}
In Eq.~\eqref{eq:series_expansion}, $\mathbb{E}_{\theta}[ \mathrm{Tr}(\hat{O}\rho(t))]$ denotes the expectation value of the desired operator in the presence of noise that is accessible to us in experimental measurements, whereas $\mathrm{Tr}(\hat{O}\rho_{\theta=0}(t))$ is the noise-free expectation value that we aim to recover. The rest of the terms in the first row of Eq.~\eqref{eq:series_expansion} contain a polynomial function in $\delta$ that can be extracted via a polynomial extrapolation (e.g., the Richardson extrapolation); in contrast, the second row describes the error term that is not recoverable via a polynomial extrapolation, which we define as the remainder term $R_{r+1}(\mathcal{D}_{\theta},t,\hat{V})$ in our zero-noise extrapolation protocol:
\begin{align}
R_{r+1}(\mathcal{D}_{\theta},t,\hat{V} ) \equiv \mathbb{E}_{\theta}\left[
\delta^{r+1}(-i)^{r+1}\int_0^t \dots \int_0^{t_{r}}\mathrm{d}t_{r+1}\dots \mathrm{d}t_1 \mathrm{Tr}\left(\hat{O}_{\mathrm{I}}(t) [\hat{V}_\mathrm{I}(t_1),[ \dots , [\hat{V}_\mathrm{I}(t_{r+1}), \rho_\mathrm{I}(t_{r+1})]\dots]]\right)\right]
. 
\end{align}
Our goal now is to bound the remainder term. Using Cauchy's mean value theorem---there exists a value of $c$ such that $\int_0^t f(t_1)\mathrm{d}t_1 = tf(c)$---there exist  $c_{1}, \dots ,c_{r+1}$ such that 
\begin{align}R_{r+1}(\mathcal{D}_{\theta},t,\hat{V} ) = \mathbb{E}_{\theta}\left[ 
    \frac{\delta^{r+1}t^{r+1}}{(r+1)!}
    \mathrm{Tr}\left(\hat{O}_{\mathrm{I}}(t) [\hat{V}_\mathrm{I}(c_1),[ \dots , [\hat{V}_\mathrm{I}(c_{r+1}), \rho_\mathrm{I}(c_{r+1})]\dots]]\right) 
    \right]
    .
\end{align}
This expression can be simplified using $f(\delta) = \frac{\delta^{r+1}t^{r+1}}{(r+1)!}$ and $g(\delta) = \mathrm{Tr}\left(\hat{O}_{\mathrm{I}}(t) [\hat{V}_\mathrm{I}(c_1),[ \dots , [\hat{V}_\mathrm{I}(c_{r+1}), \rho_\mathrm{I}(c_{r+1})]\dots]]\right) $,  where in the term $g(\delta)$, $\rho_I (c_{r+1})$ implicitly depends on $\delta$. Taking the absolute value of the remainder gives $|R| = |\mathbb{E}_\theta[f(\delta)g(\delta)]|$, which is further bounded by 
\begin{align}
    |\mathbb{E}_{\theta}[f(\delta)g(\delta)]| &\leq \mathbb{E}_{\theta}[|f(\delta)g(\delta)|]  \\ &\leq \mathbb{E}_{\theta}[|f(\delta)||g(\delta)|].\label{eq:fg-bound}
\end{align}

An application of von Neumann's inequality followed by Hölder's inequality for Euclidean spaces give a bound on the absolute value of the trace $|\mathrm{Tr}(\hat A^{\dagger}\hat B)| \leq \|\hat A\|_p \|  \hat{B} \|_q$, for $1/p + 1/q = 1$ and $\|\hat A\|_p$ the Schatten p-norm. Applying this inequality to $|g(\delta)|$ with $p=\infty$ and $q=1$ yields
\begin{align}    
|g(\delta)| &=  \left|\mathrm{Tr}\left(\hat{O}_{\mathrm{I}}(t) [\hat{V}_\mathrm{I}(c_1),[ \dots , [\hat{V}_\mathrm{I}(c_{r+1}), \rho_\mathrm{I}(c_{r+1})]\dots]]\right)\right|\\
&\leq \|\hat O_I(t) \|_\infty \left\| [\hat{V}_\mathrm{I}(c_1),[ \dots , [\hat{V}_\mathrm{I}(c_{r+1}), \rho_\mathrm{I}(c_{r+1})]\dots]]\right\|_{1}.
\end{align}
The Schatten p-norms are invariant under unitary transformations, so $\|\hat{O}_{\mathrm{I}}(T)\|_{p} = \|\hat{O}\|_{p}$, and we can further bound the nested commutator by recursively applying a bound from the  the triangle inequality and another application of the Hölder inequality:
\begin{align}
   \|[\hat{A},\hat{B}]\|_1 &= \|\hat{A}\hat{B}-\hat{B}\hat{A}\|_1\\
   &\leq 2\|\hat{A}\hat{B}\|_1 \\
   &\leq 2\|\hat A\|_\infty \|B\|_1,
\end{align}
to find the bound on $|g(\delta)|$:
\begin{align}
    |g(\delta)| \leq \|\hat O\|_{\infty}2^{r+1}\|\hat V\|^{r+1}_\infty \|\hat \rho\|_1.
\end{align}
This can then be simplified with $\|\hat \rho \|_1 = 1$. This bound is now independent of $\delta$, so combining it with Eq.~\eqref{eq:fg-bound} leads to the following bound:
\begin{align}
    |R_{r+1}(\mathcal{D}_{\theta},t,\hat{V} )| 
\leq&~
\mathbb{E}_{\theta}\left[ \left| \frac{\delta^{r+1}t^{r+1}}{(r+1)!}\right|\right] \|\hat O\|_{\infty}2^{r+1}\|\hat V\|_\infty^{r+1} \\
=&~\|\hat O\|_{\infty}\frac{(2t)^{r+1}\mathbb{E}_{\theta}\left[ \left| \delta^{r+1}\right|\right] \|\hat V\|_\infty^{r+1}}{(r+1)!}.\label{eq:full-remainder}
\end{align}

We see that the zero-noise limit expectation value can be expressed as a series in the moments of the distribution $\mathcal{D}_{\theta}$ up to some bounded remainder term. We now define measured expectations of $\hat{O}$ on a device with finite shots $N$  as $\tilde{E}_{\theta_{i},T} = \mathrm{Tr}(\hat{O}\rho_{\theta_i}(t)) + \xi_i$, where, in the large sample limit $\xi_i$ follows a normal distribution with variance
$$\mathrm{Var}(\tilde{E}_{\theta_i,t})=\nu^2_i= \frac{\left(  \mathrm{Tr(\hat{O}^{2}\rho_{\theta_{i}}(t))}-\mathrm{Tr(\hat{O}\rho_{\theta_{i}}(t))^{2}}\right)}{N}.$$
Given a set of $r+1$ measurements over a range of values $\{(\theta_0, \tilde{E}_{\theta_0,t}),\ldots, (\theta_r, \tilde{E}_{\theta_r,t})\}$, we will now apply Richardson extrapolation and bound the resulting zero-noise limit estimator in the $\theta \to 0$ limit. 

First, we must make an assumption about the distribution $\mathcal{D}_{\theta}$---that is, the moments $\mathbb{E}_{\theta}[\delta^k]$ can be expressed as a polynomial $p_k(\theta)$ in the distribution parameter $\theta$ with $\mathrm{deg}(p_k(\theta)) \leq k$. This property follows from our definition of a mitigable distribution, Def.~\ref{lemma:mitigable-distributions}. Our Dyson series expansion then is
\begin{align}
    \mathbb{E}_{\theta}\left[ \mathrm{Tr}\left(\hat{O}\rho(t)\right)\right]
    =
    \mathrm{Tr}\left(\hat{O}\rho_{\theta=0}(t)\right)+  \sum_{k=1}^{r} p_k(\theta)  a_k+R_{r+1}(\mathcal{D}_{\theta},t,\hat{V} ) , 
\end{align}
which can be rearranged as as a polynomial in $\theta$ as
\begin{align}
    \mathbb{E}_{\theta}\left[  \mathrm{Tr}\left(\hat{O}\rho_{\delta}(t)\right)\right]
    &=
    \mathrm{Tr}\left(\hat{O}\rho_{\theta=0}(t)\right)+  \sum_{k=1}^{r} p_k(\theta)  a_k + R_{r+1}(\mathcal{D}_{\theta},t,\hat{V} ) \\
    &= 
    \mathrm{Tr}\left(\hat{O}\rho_{\theta=0}(t)\right)+  \sum_{k=1}^{r} \sum_{j=0}^{k} p_{k,j}\theta^{j}  a_k  +R_{r+1}(\mathcal{D}_{\theta},t,\hat{V} )\\
    &= \mathrm{Tr}\left(\hat{O}\rho_{\theta=0}(T)\right)+ \sum_{j=0}^{r}\theta^{j}\left(\sum_{k=j}^{r}p_{k,j}a_{k}\right) +R_{r+1}(\mathcal{D}_{\theta},t,\hat{V} ) 
    .
\end{align}
The above equation is a power series in $\theta$ as desired. We note that there cannot be a $\theta^0$ contribution in the polynomial $p_k(\theta)$, for any $k$, outside of contributions from a non-zero average independent of $\theta$, as the noise should identically converge to zero when $\theta =0$. We are left with a power series 
\begin{equation}
    \mathbb{E}_{\theta}\left[ \mathrm{Tr}\left(\hat{O}\rho(t)\right)\right]  =  \mathrm{Tr}\left(\hat{O}\rho_{\theta=0}(t)\right)+ \sum_{j=1}^{r}\theta^{j}\left(\sum_{k=j}^{r}p_{k,j}a_{k}\right) +R_{r+1}(\mathcal{D}_{\theta},t,\hat{V} ), \label{eq:final-power-series}
\end{equation}
which serves as the basis of the polynomial extrapolation step in our protocol. 

Referencing~\cite{temme2017, Krebsbach2022}, 
the Richardson extrapolation estimator can be expressed in terms of the measured expectations and the Lagrange basis polynomials evaluated at $\theta=0$, $\gamma_i = \Pi_{i \neq j} \frac{\theta_j}{\theta_j - \theta_i}$, as follows: 
\begin{align*}
    \tilde{E}_0 = \sum_{i=0}^{r}\gamma_i \tilde{E}_{\theta_i, T}.
\end{align*}
For the derivations that follow, it is useful to note that the Lagrange basis polynomials satisfy the following equalities, $\sum_{i=0}^{r}\gamma_{i}=1$ and $\sum_{i=0}^{r}\gamma_{i}\theta^{k} _{i}=0$. We aim to bound the error of the estimator in terms of the Dyson series truncation error and projection noise $\xi_i$. 

Each measured expectation value is expressed as $\tilde{E}_{\theta_i, T} = E_{0}+\sum_{k=1}^{r} a_{k}\theta_{i}^{k}+R_{r+1}(\mathcal{D}_{\theta_i},T,\hat{V}) + \xi_{i}$. Substituting the series expansion for the measured expectation values into the Richardson estimator gives
\begin{align}
\tilde{E}_{0,t}&= \sum_{i=0}^{r}\gamma_{i}\tilde{E}_{\theta_{i},t} \\
&= \sum_{i=0}^{r}\gamma_{i}\left( E_0 + \sum_{k=1}^{r}a_{k}\theta_{i}^{k} +R_{r+1}(\mathcal{D}_{\theta_i},t,\hat{V}) + \xi_{i}\right)  \\
&= \sum_{i=0}^{r}\gamma_{i}E_0 +   \sum_{k=1}^{r}a_{k}\left(  \sum_{i=0}^{r}\gamma_{i}\theta_{i}^{k} \right) + \sum_{i=0}^r \gamma_{i}\left(   R_{r+1}(\mathcal{D}_{\theta_i},t,\hat{V}) + \xi_{i}\right) \\
&= \left(   \sum_{i=0}^{r}\gamma_{i}\right)E_0 +   \sum_{k=1}^{r}a_{k}\left(  \sum_{i=0}^{r}\gamma_{i}\theta_{i}^{k} \right) + \sum_{i=0}^r \gamma_{i}\left(   R_{r+1}(\mathcal{D}_{\theta_i},t,\hat{V}) + \xi_{i}\right) \\
&= E_{0} + \sum_{i=0}^r \gamma_{i}\left(   R_{r+1}(\mathcal{D}_{\theta_i},t,\hat{V}) + \xi_{i}\right).
\end{align}

Now, given the estimator variance $\nu_i^2$, we can bound the error from sampling as follows.
Letting $|\mathrm{Tr}(\hat O \hat \rho_{\theta_i}(t))-\tilde E_{\theta_i, t}| = |\xi_i|$ with $\xi_i \sim \mathcal{N}\left( 0, \nu_i^2 \right)$, with probability $1-\epsilon$, $|\xi_i| \leq \sqrt{ 2 }\nu_i\mathrm{erf}^{-1}(1-\epsilon)$, where $\mathrm{erf}^{-1}(\cdot)$ is the inverse error function. Thus, we can bound the error in the estimator probabilistically according to the distributions generating $\xi_{i}$. Assuming, for each $\xi_{i}$, the variance is $\nu_{i}\approx \nu$, with probability $1-\epsilon$, 
\begin{align}
|\tilde{E}_{0,T}-E_{0}| \leq \sum_{i=0}^{r}|\gamma_{i}|\left| R_{r+1}( \mathcal{D}_{\theta_i},T,\hat{V}) + \sqrt{ 2 }\nu \mathrm{erf}^{-1}(1-\epsilon) \right|.
\end{align}
\qed

We see that it is sufficient to only demonstrate that the moments $\mathbb{E}_{\theta_i}[\delta^k]$ for $k  \leq r$ are polynomials in $\theta$, and that the $(r+1)^\mathrm{th}$ absolute moment is bounded. 

\subsection{Normal Distribution}\label{sec:normal-bounds}
\begin{corollary} [Analog ZNE for Gaussian-distributed
noise]
    The zero-centered normal distribution defined in Eq.~\eqref{eq:Gauss_def} is mitigable and permits ZNE along the $\sigma^2$ axis with the remainder of Richardson extrapolation on $r+1$ experimental estimators given by \begin{equation}|R_{r+1} (\mathcal{D}_{\sigma^2},t, \hat V)|\leq \|\hat{O}\|_\infty\frac{\left(\sigma^{2}(2t)^2\|\hat{V}\|_\infty^2\right)^{r+1}}{(2r+2)!!}.\end{equation}
\end{corollary}

\proof
For a zero-centered normal distribution $\mathcal{D}_{\sigma^2}  = \mathcal{N}(0, \sigma^2)$, the moments can be computed to be 
\begin{align}
    \mathbb{E}[\delta ^ k] = 
    \begin{cases}
    \sigma^k (k-1)!! &\text{for }\, k\, \text{ even}, \\ 0 &\text{for}\, k\, \text{odd},
    \end{cases}
\end{align}
where $(k-1)!!$ is the double factorial. The absolute moments are 
\begin{align}
\mathbb{E}[|\delta|^k] = \sigma^k 2^{k/2} \frac{\Gamma(\frac{k+1}{2})}{\sqrt{\pi}} 
.
\end{align}

The zero-centered normal distribution exhibits unique moment properties, leading to a tailored extrapolation approach. As the odd raw moments are zero and the even moments are nonzero, the Dyson series expansion only includes even terms, see Eq.~\eqref{eq:series_expansion}. Additionally, the expansion's dependence on even powers of the standard deviation, $\sigma$, favors extrapolation with respect to $\sigma^2$ rather than $\sigma$. This effectively doubles the order of the polynomial fit computed using Richardson extrapolation, corresponding to the remainder term $R_{2(r+1)}$ from Eq.~\eqref{eq:final-power-series} using a dataset of $r+1$ estimators:
\begin{align}
    |R_{2(r+1)} (\mathcal{D}_{\theta_i},t,\hat{V})|\leq \|\hat{O}\|_{\infty}\frac{\left(\|\hat{V}\|_\infty^2\sigma^{2}(2t)^2\right)^{r+1}}{(2r+2)!!}.
\end{align}
The series has an infinite radius of convergence as the remainder always trends towards zero due to the $(r+1)!!$ term in the denominator:
\begin{align}
    \lim_{r \rightarrow \infty} |R_{2(r+1)}(\mathcal{D}_\theta, t,\hat{V})| = 0.
\end{align}
\qed
\subsection{Thermal Distribution}\label{sec:thermal-bounds}

\begin{corollary}[ZNE for thermally distributed noise] 
Thermal noise is mitigable for times $t < \left(2(\bar{n}+1)\|\hat{V}\|_\infty\right)^{-1}$ and permits zero-noise extrapolation along the $\theta = \bar{n}$ axis with remainder \begin{equation}
        \left|R_{r+1}(\bar{n}, t, \hat V)\right| \leq \|\hat{O}\|_\infty\left(2(\bar{n}+1)t\|\hat{V}\|_\infty \right)^{r+1}.
    \end{equation} 
\end{corollary}

\proof
Our goal is to derive and bound the moments of the probability distribution $\mathrm{Pr}(n)$ for occupying Fock state $\ket{n}$ of an harmonic oscillator mode. This distribution, for a system with Hamiltonian $\hat H=\epsilon \hat n$ at inverse temperature $\beta$, is derived directly from our microscopic model of center-of-mass phonon heating. 

\subsubsection{Defining the Thermal distribution}
Starting from basic definitions for the Bose-Einstein distribution, we have
\begin{align}
    \mathcal{Z} &= \frac{1}{1-e^{-(\varepsilon-\mu)\beta}}, \\
    \bar{n} &= \frac{1}{\beta\mathcal{Z}}\left(\frac{\partial \mathcal{Z}}{\partial \mu}\right)_{V,\beta} = \frac{1}{e^{(\varepsilon-\mu)\beta}-1}, 
\end{align}
which yields the thermal Bose distribution for $\mu=0$:
\begin{align}
    \mathrm{Pr}(n) &= \frac{e^{-\varepsilon\beta n}}{\mathcal{Z}} = \left(1-e^{-\varepsilon\beta}\right)e^{-\varepsilon\beta n}.
\end{align}
We can express $\mathrm{Pr}(n)$ using only $\bar{n}$ and $n$ by substituting 
\begin{align}
    \left(1-e^{-\varepsilon\beta}\right) &= \frac{1}{\bar{n}+1}, \quad
    e^{-\varepsilon\beta } = \frac{\bar{n}}{\bar{n}+1} ,
\end{align}
to give 

\begin{equation}
    \mathrm{Pr}(n)=\frac{1}{\bar{n}+1}\left( \frac{\bar{n}}{\bar{n}+1} \right)^{n}.
\end{equation}
This is a geometric distribution of the form $\mathrm{Pr}(n) = p(1-p)^n$ with 
\begin{equation}
p=\frac{1}{\bar{n}+1}.
\label{eq:p-geometric-vs-n-thermal}
\end{equation} 
\subsubsection{The Thermal Distribution is Mitigable}
Making use of the moment generating function, the following Lemma introduces a mechanism to determine if the moments $\mathbb{E}_\theta[\delta^k]$ of a distribution are polynomials in the distribution parameter $\theta$ with degree at most $k$. This condition is sufficient for a distribution to be mitigable, and allows us to make use of established results on the geometric distribution in proving results for the thermal distribution. 
\begin{lemma}[Mitigable Distribution]
A distribution $\mathcal{D}_{\theta}$ with moment generating function $f(s, \theta) \coloneqq  \mathbb{E}_{\mathcal{D}_{\theta}(\delta)}[e ^{s \delta}]$ 
is mitigable in $\theta$ if, for all $j$, $b_j(t)$ in the Taylor expansion of $f(s, \theta) = \sum_{j=1}^\infty \frac{b_{j}(s)}{j!}\theta^{j}$ have zeros of order $j$ at $s=0$. 
\end{lemma}

\proof
\noindent{The $k^\mathrm{th}$ moment of $\mathcal{D}_{\theta}$ is given by }$\frac{\partial^{k}}{\partial s^k}f(s,\theta)|_{s=0}$. We now Taylor expand $f(s,\theta)$ with respect to $\theta$ to get $f(s,\theta) = \sum_{j} \frac{b_{j}(s)}{j!} \theta^j$. Assuming the coefficient $b_{j}(s)$ has a zero of order $j$ at $s=0$, the derivative $\sum_{j}\frac{\partial^{k}}{\partial s^{k}}\frac{b_{j}(s)}{j!} \theta^{j}|_{s=0 }= \sum_{j=1}^{k} \frac{b^{(k)}_{j}(0)}{j!}\theta^j$ is a polynomial of degree at most $k$ in $\theta$, where we have used $\frac{\partial^{k} b_{j}(s)}{\partial s^{k}} = b_{j}^{(k)}(s)$. \qed

The moment generating function $f(s,p )$ for the geometric distribution is 
\begin{equation}
    \tilde{f}(s, p) = \frac{p}{(1-(1-p)e^s)},
\end{equation}
into which we substitute Eq.~\eqref{eq:p-geometric-vs-n-thermal} to give an expression in terms of $\bar{n}$:
\begin{align}
    \tilde{f}(s, \frac{1}{\bar{n}+1}) 
    &= \frac{\frac{1}{\bar{n}+1}}{(1-(1-\frac{1}{\bar{n}+1})e^s)} \\
    &= \frac{1}{(1-e^s)\bar{n}+1} \\
    &= f(s,\bar{n}).
\end{align}

We compute the Taylor series for $f(s, \bar{n})$, expanded around $\bar{n}=0$, by first computing the derivatives
\begin{align}
    \frac{\partial^k}{\partial \bar{n}^k} \frac{1}{(1-e^s)\bar{n}+1}  &= \frac{(-1)^{k}k!(1-e^s)^{k}}{((1-e^s)\bar{n}+1)^{k+1}}
\end{align}
and writing out the series expansion
\begin{equation}
    f(s,\bar{n}) = \sum_{k=0} ^{\infty} \frac{(-1)^k (1-e^s)^k}{((1-e^s)\bar{n}+1)^{k+1}}\bar{n}^k.
\end{equation}
It is now clear that each coefficient has a zero of order $k$ at $s=0$, and the moments $\mathbb{E}[ n^k ]$ are polynomial in $\bar n$, which we denote as
\begin{align}
\label{seq:therm.n.moments}
\mathbb{E}[ n^{k} ] = \sum_{i=0}^k p_i \bar{n}^i 
.
\end{align}

\subsubsection{Bounding the Remainder}
We have shown that the thermal distribution is mitigable. What remains is to bound the remainder Eq.~\eqref{eq:full-remainder} for the thermal distribution. Again, we will make use of known results on the equivalent geometric distribution and substitute Eq.~\eqref{eq:p-geometric-vs-n-thermal}. In particular, the moments of the geometric distribution can be expressed using the polylogarithm function $\mathrm{Li}_k(z)$. We use the definition for $\mathrm{Li}_{-k}(z)$ for negative integers $-k$ using a sum over Eulerian numbers,
\begin{equation}
    \mathrm{Li}_{-k}(z) = \frac{1}{(1-z)^{k+1}} \sum_{j=0}^{k-1} \left\langle \genfrac{}{}{0pt}{}{k}{j} \right\rangle z^{k-j},
\end{equation}
from which we will primarily make use of the property $\sum_{j=0}^{k-1}\left\langle  \genfrac{}{}{0pt}{}{k}{j}   \right\rangle = k!$ in our bounds on computing the moment to order $k$. The $k^\mathrm{th}$ moment then is expressed as
\begin{align}
    \mathbb{E}[ n^k ] &=p\mathrm{Li}_{-k}(1-p) = \frac{p}{p^{k+1}}\sum_{j=0}^{k-1}\left\langle   \genfrac{}{}{0pt}{}{k}{j}\right\rangle (1-p)^{k-j}= \frac{1}{p^{k}}\sum_{j=0}^{k-1}\left\langle   \genfrac{}{}{0pt}{}{k}{j}\right\rangle (1-p)^{k-j}.
\end{align}
Substituting the definitions of $p$ from Eq.~\eqref{eq:Prdef} gives
\begin{align}
\mathbb{E}[ n^k ] &= {(\bar{n}+1)^{k}} \sum_{j=0}^{k-1}\left\langle   \genfrac{}{}{0pt}{}{k}{j}\right\rangle \frac{\bar{n}^{k-j}}{(\bar{n}+1)^{k-j}},  
\end{align} 
which we can bound using $\bar{n} <  (\bar{n}+1)$ and the sum over the Eulerian numbers:
\begin{align}
\mathbb{E}[ n^k ] &= {(\bar{n}+1)^{k}} \sum_{j=0}^{k-1}\left\langle   \genfrac{}{}{0pt}{}{k}{j}\right\rangle \frac{\bar{n}^{k-j}}{(\bar{n}+1)^{k-j}}  \\ 
&< {(\bar{n}+1)^{k}} \sum_{j=0}^{k-1}\left\langle   \genfrac{}{}{0pt}{}{k}{j}\right\rangle \frac{(\bar{n}+1)^{k-j}}{(\bar{n}+1)^{k-j}} \\
&= {(\bar{n}+1)^{k}} \sum_{j=0}^{k-1}\left\langle   \genfrac{}{}{0pt}{}{k}{j}\right\rangle \\
&= (\bar{n}+1)^{k}k!.
\end{align} 
Thus, we have our bound on the error term as $|R_{r+1}(\mathcal{D}_{\bar{n}},t,\hat{V})| < \|\hat{O}\|_\infty\|\hat{V}\|_\infty^{r+1}(2t(\bar{n}+1))^{r+1}$. \qed

This means $\lim_{r\to\infty}|R_r(\mathcal{D}_{\bar{n}},t,\hat{V})|=0$ for $t \leq (2(\bar{n}+1)\|\hat V\|_\infty)^{-1} $, which provides a finite lower bound on the radius of convergence for the expansion Eq.~\eqref{eq:final-power-series}. This is in contrast to our result on the normal distribution, which we demonstrate to have an infinite radius of convergence with respect to $t$.

\subsection{Central Moment Bound}
As we demonstrate in Appendix~\ref{app:rabi-correction}, the frequency calibration step we perform for the thermal distribution shifts the distribution from $\mathcal{D} (\delta)$ to $\mathcal{D}(\delta - \alpha \bar n)$. This corresponds to computing the central moment, instead of the raw moment, of the distribution. In the section which follows, we compute a bound on the even central moments to demonstrate that the upper bound on the radius of convergence is increased by a factor which converges to $e$ in the $r\to \infty$ limit. The $k^\mathrm{th}$ central moment of the thermal distribution is

\begin{align}
    \mathbb{E}_{\alpha\bar n} \left[ \alpha^k(n - \bar n)^k\right] = \alpha^k\sum_{j=0}^\infty \frac{(n-\bar n)^k}{\bar n + 1} \left(1-\frac{1}{\bar n+1}\right)^n.
\end{align}
The probability mass function is bounded by the exponential function,
\begin{align}
    \frac{1}{\bar n + 1}\left(1-\frac{1}{\bar n+1}\right)^n < \frac{e^{\frac{-n}{\bar n + 1}}}{\bar n + 1}
\end{align}
and the summation correspond to the right Riemann sum of the probability mass function from $-1$ to $\infty$. As the probability mass function is monotonic decreasing, the integral upper bounds the summation. These are sufficient to bound the central moment:
\begin{align}
    \mathbb{E}_{\alpha\bar n} \left[ \alpha^k(n - \bar n)^k\right] = \alpha^k\sum_{j=0}^\infty \frac{(n-\bar n)^k}{\bar n + 1} \left(1-\frac{1}{\bar n+1}\right)^n &< \alpha^k \int_{-1}^\infty\,dn\, \frac{(n-\bar n)^k}{\bar n + 1} \left(1-\frac{1}{\bar n+1}\right)^n \\
    &< \alpha^k \int_{-1}^\infty\,dn\, \frac{(n-\bar n)^k}{\bar n + 1} e^{\frac{-n}{\bar n + 1}} \\
    &= e^{\frac{-1}{\bar n +1}} (\bar n + 1)^k k! \sum_{j=0}^k \frac{(-1)^j}{j!}.
\end{align}
For large $\bar n$ and $k$ this upper bound converges to $ {k! (\bar n + 1)^k}/{e}$, or a factor of $1/e$ improvement in the bound than compared to the raw moments. For small values of $k$, as typically considered for least-squares fitting, the improvement on the bound is between to $1/3$ and $1/2$. 

\section{Trapped ion Rabi Frequency Calibration}\label{app:rabi-correction}
In this section, we elaborate on the effect of the single-qubit frequency calibration and generalize the result to two-body interactions. While we again emphasize that this process is not required for our ZNE method, we observe significant improvements in the results of the extrapolation with the calibration procedure implemented, as demonstrated in 
\begin{figure}
    \centering
    \includegraphics[width=\linewidth]{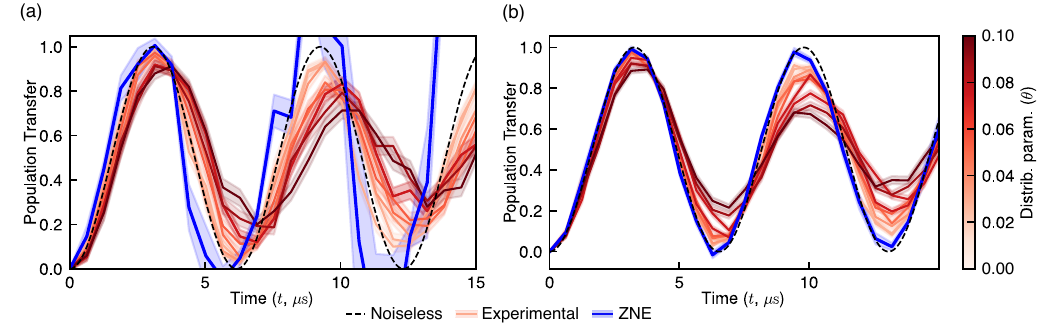}
    \caption{\textbf{Frequency calibration ZNE comparison.} Data from Fig.~\ref{fig:3}, showing single-qubit Rabi oscillations. Red lines (shaded region) are experimental estimators (error of the mean) with color determined by the distribution parameter $\theta$, where darker shades of red are more noisy; black dashed lines are the noise-free results; and blue lines (shaded region) are the result of ZNE using LOOCV least-squares fitting (polynomial error). (a) Rabi oscillations without any frequency calibration, and the ZNE result. (b) Rabi oscillations with frequency calibration and the first order term in the least-squares fit polynomial set to zero.}
    \label{fig:sup_fig_3}
\end{figure}
Fig.~\ref{fig:sup_fig_3}, which shows the result of the extrapolation for the non-calibrated data from Fig.~\ref{fig:3}(a) compared to extrapolation after calibration, Fig.~\ref{fig:3}(c).

The frequency compensation procedure aims to center the thermal distribution to correct for the first-order effect of the non-zero average frequency for $\bar n \neq 0$, that is, transform the distribution $\delta \sim \mathcal{D}_{\alpha\bar n} \to \delta-\bar n$. Considering the derivation in Ref.~\cite{cetina2022} of the single-qubit thermal Rabi oscillations, we aim to shift the integral over the thermal distribution:
\begin{align}
    \int e^{-\alpha \beta \omega n} \sin^{2}(\Omega t(1+\alpha n))\, \mathrm{d}n 
\to& \int e^{-\alpha \beta \omega n} \sin^{2}(\Omega t(1+\alpha (n-\bar{n})))\, dn \\
=& \frac{1-C \cos(\Omega t - \phi(t) -\alpha \bar{n}\Omega t)}{2}.
\end{align}
Thus, shifting the distribution corresponds to changing the input frequency according to $\Omega \to \Omega(1-\alpha \bar n)$. For 2-body interactions, this same frequency shift corrects the linear-order error. Consider the short-time expansion $\mathbb{E}_{\alpha \bar{n}}\left[e^{-i\hat Ht}\approx 1-i\hat H t +\mathcal{O}(t^{2})\right]$, with $\hat{H}= \sum_{ij} \Omega_{i}(1+\alpha_{i} n)\Omega_{j}(1+\alpha_{j} n)\gamma_{ij}\hat{\sigma}^z_{i}\hat{\sigma}^z_{j}$ and $\gamma_{ij}$ depending on the phonon modes and other experimental parameters. Performing the replacement $\Omega_{i}\to \tilde{\Omega}_{i}=\Omega_i(1-\alpha_{i}\bar{n})$ and considering the coefficient of the two-body terms gives
\begin{align}
&\mathbb{E}_{\alpha\bar{n}}[\tilde{\Omega}_{i}(1+\alpha_{i} n)\tilde{\Omega}_{j}(1+\alpha_{j} n)\gamma_{ij}] 
\nonumber \\
= & \gamma_{ij}\Omega_{i}\Omega_{j}(1+ \alpha_{i}\mathbb{E}_{\alpha \bar{n}}[n] + \alpha_{j}\mathbb{E}_{\alpha \bar{n}}[n] + \alpha_{i}\alpha_{j}\mathbb{E}_{\alpha \bar{n}}[n^{2}]) 
(1- \alpha_{i}\bar{n} - \alpha_{j}\bar{n} + \alpha_{i}\alpha_{j}\bar{n}^{2}) \nonumber\\
=&\Omega_{i}\Omega_{j}(1 + \alpha_{i}\bar{n} + \alpha_{i} \bar{n} + \alpha_{j}\alpha_{j}\mathbb{E}[n^{2}]- \alpha_{i}\bar{n} - \alpha_{j}\bar{n} + \alpha_{i}\alpha_{j}\bar{n}^{2}   -\alpha_{i}^{2} \bar{n}^{2} - \alpha_{j}^{2}\bar{n}^{2} - 2\alpha_{i}\alpha_{j}\bar{n}^{2}+ \alpha_{i}\alpha_{j} \bar{n}^{2} \mathcal{O}(\alpha^{3}\bar{n}^{3})) \nonumber\\
=& \Omega_{i}\Omega_{j}(1 + \mathcal{O}(\alpha^{2} \bar{n}^{2})).
\end{align}
We have neglected terms $\alpha^{k} \bar{n}^{l}$ with $l<k$ as $\alpha \ll 1$. For both the single-body and two-body terms in the Hamiltonian, the single-qubit frequency calibrations approximately correct the effects of the frequency fluctuations induced by thermal motion that are linear in the average phonon occupation $\bar{n}$. 

\section{Experimental Apparatus}\label{app:experiment}
In this section, we introduce our trapped-ion apparatus and provide details about the experimental parameters used in section~\ref{sct:ions}.

We trap a chain of $N=27$ $^{171}\text{Yb}^{+}$ ions in a surface ion trap held in a single potential well \cite{Feng2023,katz2024observing}. The spin states are encoded in the hyperfine levels of the electronic ground state $\ket{\downarrow}=\ket{F=0, m_F=0}$ and $\ket{\uparrow}=\ket{F=1, m_F=0}$, separated by an energy $\tilde{\omega}_0=2\pi\times12.6\mathrm{GHz}$. A pair of 355 nm laser beams drives a Raman transition between these levels. One beam provides a nearly uniform global intensity across the ion chain, while an array of tightly focused individual addressing beams (propagating perpendicular to the other beam) is controlled by a  32-channel acousto-optic modulator (AOM). The Raman beatnote between the two beams can drive single-qubit rotations of the form  $\hat \sigma_\phi=\cos\phi\hat{\sigma}^x +\sin\phi\hat{\sigma}^y$ when its beatnote frequency matches $\tilde{\omega}_0$. The azimuthal angle $\phi$ is set by the beatnote phase. 

Our ion crystal supports 81 phonon modes representing the three dimensional motion of the ion chain. A set of 27 modes corresponds to collective \textit{axial} motion, with nonzero phonon-distribution of the lowest-frequency modes (e.g., the COM mode) being the primary source of error studied in this work; these errors are associated with the second term in 
Eq.~(\ref{eq:fock-state-omega}). The remaining \textit{radial} modes mediate the Ising coupling between the spins. We align the wavevector difference between the global and individual beams, $\delta \vec k$, along one of the radial principal axes of the trapping potential, ensuring that the Ising interaction is primarily mediated by a single set of $27$ radial phonon modes. The frequencies of these modes, $\omega_n$ ($1\leq n\leq27)$, range from $\omega_1=2\pi\times2.74\mathrm{MHz}$ (the lowest frequency zig-zag mode in this radial direction) to $\omega_{27}=2\pi\times 3.05\mathrm{MHz}$ (the highest-frequency COM mode). A spin-spin interaction described by Eq.~\eqref{eq:two-qubit-ham}
is driven using Raman beams with two simultaneous beatnote frequencies $\omega_{\pm}=\tilde{\omega}_0\pm \delta$. In this experiment, we set  $0<\delta<\omega_1$. We operate in the far-detuned regime suitable for quantum simulation, where mode excitations are weak ~\cite{monroe2021}. In this regime, the interaction strength of the Ising Hamiltonian is given by  $J\propto\Omega^{i}\Omega^j/\delta$, and the noise described in the main text dominates the simulation.

At the start of each experimental cycle, the ions undergo Doppler cooling, followed by resolved sideband cooling to bring the radial modes to their motional ground state. The axial modes, however, are only Doppler cooled, as the Raman beams used for sideband cooling are perpendicular to the axial direction and cannot efficiently cool these modes. Due to anomalous heating, the ions heat up in the axial direction while the radial modes are cooled. The ions are then optically pumped into state $\ket{\downarrow}$. Next, a delay of duration $t_{\mathrm{w}}\geq0 $ is applied, during which anomalous heating causes the ion motion to heat up, predominantly affecting the long-wavelength, low-frequency modes of the chain in the axial direction \cite{cetina2022,katz2023programmable}. This process provides a controllable means of increasing the effective temperature of initial phonon distribution, amplifying the noise.
For the 2-qubit experiment in Sec.~\ref{sct:exp_twoq}, a pair of neighboring ions, indexed $i$ and $i+1$, are initialized in the $\ket{\uparrow \downarrow}$ state via a $\pi$-pulse $\hat{\sigma}^x_{i}$ rotation. After the variable wait time, the Hamiltonian in Eq.~\eqref{eq:two-qubit-ham} is applied for the target evolution time $t$. The final state is measured using state-dependent fluorescence detection.
\end{widetext}
\end{document}